\documentclass[twocolumn]{revtex4}

\usepackage{graphicx}
\usepackage{listings}
\usepackage{slashed}
\usepackage{MnSymbol}
\newcommand{\ac}[1]{#1}

\newcommand{\acs}[1]{#1}
\newcommand{\acp}[1]{#1}

\newcommand{\texorpdfstring}[2]{#1}
\newcommand{\burl}[1]{\url{#1}}

\newcommand{\be}{\begin{equation}}
\newcommand{\ee}{\end{equation}}
\newcommand{\ba}{\begin{eqnarray}}
\newcommand{\ea}{\end{eqnarray}}

\newcommand{\pderiv}[2]{\frac{\partial #1}{\partial #2}}

\newcommand{\bra}[1]{\langle #1|}
\newcommand{\ket}[1]{|#1\rangle}
\newcommand{\braket}[2]{\langle #1|#2\rangle}

\renewcommand{\vec}[1]{ {\bf #1}}

\def\ie{{\em i.e.}\ }

\newcommand{\Aphi}{A_\phi(\rho)}

\newcommand{\bhattext}[1]{${\bf \hat{#1}}$}

\newcommand{\mean}[1]{\langle#1\rangle}

\newcommand{\Paulispin}{\frac{1}{4} {\rm tr}e^{\frac{1}{2}\int_0^T d\tau \sigma_{\mu \nu}F^{\mu \nu}(x_{\rm CM}+x(\tau))}}

\def\GeV1{{\rm GeV}^{-1}}

\begin{document}
\title{Parallel Worldline Numerics: Implementation and Error Analysis} 
\author{Dan Mazur}
\email{daniel.mazur@mcgill.ca}
\affiliation{McGill High Performance Computing Centre, 
McGill University, 1100 Rue Notre-Dame Ouest, Montreal, QC H3C 1K3}
\author{Jeremy S. Heyl}
\email{heyl@phas.ubc.ca}
\affiliation{Department of Physics and Astronomy,
University of British Columbia, Vancouver BC V6T 1Z1 Canada; Canada Research Chair}
\begin{abstract}
  We give an overview of the worldline numerics technique, and
  discuss the parallel CUDA implementation of a worldline numerics algorithm.  In
  the worldline numerics technique, we wish to generate an ensemble of representative
  closed-loop particle trajectories, and use these to compute an
  approximate average value for Wilson loops. We show how this can be
  done with a specific emphasis on cylindrically symmetric magnetic
  fields.  The fine-grained, massive parallelism provided by the GPU
  architecture results in considerable speedup in computing Wilson
  loop averages.  Furthermore, we give a brief overview of uncertainty
  analysis in the worldline numerics method.  There are uncertainties from
  discretizing each loop, and from using a statistical ensemble of
  representative loops. The former can be minimized so that the latter
  dominates. However, determining the statistical uncertainties is
  complicated by two subtleties. Firstly, the distributions generated
  by the worldline ensembles are highly non-Gaussian, and so the
  standard error in the mean is not a good measure of the statistical
  uncertainty.  Secondly, because the same ensemble of worldlines is
  used to compute the Wilson loops at different values of $T$ and
  $x_\mathrm{ cm}$, the uncertainties associated with each computed
  value of the integrand are strongly correlated. We recommend a form
  of jackknife analysis which deals with both of these problems.
\end{abstract}

\maketitle

In this contribution, we will give an overview of a numerical
technique which can be used to compute the effective actions of
external field configurations.  The technique, called either worldline
numerics or the Loop Cloud Method, was first used by Gies and
Langfeld~\cite{Gies:2001zp} and has since been applied to computation
of effective actions
\cite{Gies:2001tj,Langfeld:2002vy,Gies:2005sb,Gies:2005ym,Dunne:2009zz}
and Casimir energies
\cite{Moyaerts:2003ts,Gies:2003cv,PhysRevLett.96.220401}. More
recently, the technique has also been applied to pair
production~\cite{2005PhRvD..72f5001G} and the vacuum polarization
tensor~\cite{PhysRevD.84.065035}.  Worldline numerics 
is able to compute quantum effective actions in the one-loop 
approximation to all orders in both
the coupling and in the external field, so it is well suited to
studying non-perturbative aspects of quantum field theory in strong
background fields. Moreover, the technique maintains gauge invariance
and Lorentz invariance.  The key idea of the technique is that a path
integral is approximated as the average of a finite set of $N_l$
representative closed paths (loops) through spacetime. We use a
standard Monte-Carlo procedure to generate loops which have large
contributions to the loop average.

\section{QED Effective Action on the Worldline}

Worldline numerics is built on the worldline formalism which was
initially invented by Feynman~\cite{PhysRev.80.440, PhysRev.84.108}.
Much of the recent interest in this formalism is based on the work of
Bern and Kosower, who derived it from the infinite string-tension
limit of string theory and demonstrated that it provided an efficient
means for computing amplitudes in QCD~\cite{PhysRevLett.66.1669}.
For this reason, the worldline formalism is often referred to as
`string inspired'. However, the formalism can also be obtained
straight-forwardly from first-quantized field
theory~\cite{1992NuPhB.385..145S}, which is the approach we will adopt
here.  In this formalism the degrees of freedom of the field are
represented in terms of one-dimensional path integrals over an
ensemble of closed trajectories.

\begin{widetext}
We begin with the QED effective action expressed in the proper-time 
formalism \cite{Schwinger:1951},
\be
	\label{eqn:trln}
	\mathrm{Tr~ln}\left[\frac{\slashed{p}+e\slashed{A}_\mu^0
	-m}{\slashed{p}-m}\right] = -\frac{1}{2}\int d^4x \int_0^\infty
	\frac{dT}{T}e^{-iTm^2} 
	 \times \mathrm{tr}\biggr( \bra{x}e^{iT(\slashed{p}
	+e\slashed{A}^0_\mu)^2}\ket{x} 
	- \bra{x}e^{iTp^2}\ket{x}\biggr).
\ee
To evaluate $\bra{x}e^{iT(\slashed{p}_\mu +
  e\slashed{A}_\mu)^2}\ket{x}$, we recognize that it is simply the
propagation amplitude $\braket{x,T}{x,0}$ from ordinary quantum
mechanics with $(\slashed{p}_\mu + e\slashed{A}_\mu)^2$ playing the
role of the Hamiltonian.  We therefore express this factor in its path
integral form:
\be
	\bra{x}e^{iT(\slashed{p}_\mu + e\slashed{A}_mu)^2}\ket{x} =  \mathcal{N}
	\int \mathcal{D}x_\rho(\tau) e^{-\int_0^T d\tau \left[\frac{\dot{x}^2(\tau)}{4} 
	+ i A_\rho x^\rho(\tau)\right]} 
	\times \Paulispin.
\ee
$\mathcal{N}$ is a normalization constant that we can fix by using
our renormalization condition that the fermion determinant should
vanish at zero external field:
\be
	\bra{x}e^{iTp^2}\ket{x}  =  \mathcal{N}\int \mathcal{D} 
		x_p(\tau)e^{-\int_0^T d\tau\frac{\dot{x}^2(\tau)}{4}} 
	= \int \frac{d^4p}{(2\pi T)^4}\bra{x}e^{iTp^2}\ket{p}\braket{p}{x} 
	=  \frac{1}{(4\pi T)^2},	
\ee
We may now write
\be
	\mathcal{N}\int \mathcal{D}x_\rho(\tau) e^{-\int_0^T d\tau[\frac{\dot{x}^2(\tau)}{4}+iA_\rho x^\rho(\tau)]}
	\Paulispin
	= \frac{\left\langle e^{-i\int_0^T d\tau A_\rho x^\rho(\tau)}\Paulispin\right\rangle_x}{(4\pi T)^2} ,
\ee
where
\be
	\label{eqn:meandef}
	\mean{\hat{\mathcal{O}}}_x = \frac{\int \mathcal{D}x_\rho(\tau) \hat{\mathcal{O}} 
	e^{-\int_0^T d\tau\frac{\dot{x}^2(\tau)}{4}}}{\int \mathcal{D}x_\rho(\tau)  
	e^{-\int_0^T d\tau\frac{\dot{x}^2(\tau)}{4}}}
\ee
is the weighted average of the operator $\hat{\mathcal{O}}$ over an ensemble of closed particle 
loops with a Gaussian velocity distribution.

Finally, combining all of the equations from this section results in the 
renormalized one-loop effective action for \ac{QED} on the worldline:
\be
	\label{eqn:QEDWL}
	\Gamma^{(1)}[A_\mu] = \frac{2}{(4\pi)^2}\int_0^\infty
	\frac{dT}{T^3}e^{-m^2T}\int d^4x_\mathrm{CM} \times
        \left[\left\langle e^{i\int_0^Td\tau A_\rho(x_\mathrm{CM}+x(\tau))\dot{x}^\rho(\tau)}
	\Paulispin\right\rangle _x -1\right]. 
\ee
\end{widetext}
\section{Worldline Numerics}

The averages, $\mean{\hat{\mathcal{O}}}$, defined by equation (\ref{eqn:meandef})
involve 
functional integration over every possible closed path through spacetime
which has a Gaussian velocity distribution.  
The prescription of the worldline numerics technique is to compute 
these averages approximately using a finite set of $N_l$ representative loops 
on a computer.  The worldline average is then approximated as the mean of 
an operator evaluated along each of the worldlines in the ensemble:
\be
	\mean{\hat{\mathcal{O}}[x(\tau)]} \approx 
	\frac{1}{N_l} \sum_{i=1}^{N_l} \hat{\mathcal{O}}[x_i(\tau)].
\ee

\subsection{Loop Generation} 
\label{sec:loopgen}
The velocity distribution for the loops depends on
the proper time, $T$.  However, generating a separate ensemble of loops for 
each value of $T$ would be very computationally expensive.  This problem is alleviated by generating
a single ensemble of loops, $\vec{y}(\tau)$, representing unit proper time,
and scaling those loops accordingly for different values of $T$:
\be 
\vec{x}(\tau) = \sqrt{T}\vec{y}(\tau/T) ,
\ee
\be \int_0^T d\tau \vec{\dot{x}}^2(\tau) \rightarrow \int_0^1 dt
\vec{\dot{y}}^2(t).
\ee

There is no way to treat the integrals as continuous as we generate our loop
ensembles.  Instead, we treat the integrals as sums over discrete points
along the proper-time interval $[0,T]$.  This is fundamentally different
from space-time discretization, however.  Any point on the worldline loop
may exist at any position in space, and $T$ may take on any value.  It is
important to note this distinction because the worldline method retains
Lorentz invariance while lattice techniques, in general, do not.

The challenge of loop cloud generation is in generating a discrete set
of points on a unit loop which obeys the prescribed velocity distribution.
There are a number of different algorithms for achieving this goal that have
been discussed in the literature.  Four possible algorithms are compared
and contrasted in \cite{Gies:2003cv}.  In this work, we choose a more
recently developed algorithm, dubbed ``d-loops", which was first described
in \cite{Gies:2005sb}. To generate a ``d-loop", the number of points is iteratively 
doubled, placing the new points in a Gaussian distribution between the existing neighbour points.
We quote the algorithm directly:
\begin{quote} 
    \begin{itemize} 
	  \item[(1)] Begin with one arbitrary point
		$N_0=1$, $y_{N}$.

	  \item[(2)] Create an $N_1=2$ loop, i.e., add a point $y_{N/2}$ that is
	    distributed in the heat bath of $y_N$ with
	   \begin{equation} 
		 e^{-\frac{N_1}{4} 2 (y_{N/2} -y_{N})^2}. \label{yn2}
	   \end{equation}

      \item[(3)] Iterate this procedure, creating an $N_k=2^k$ points
        per line (ppl) loop by adding $2^{k-1}$ points
        $y_{{qN}/{N_k}}$, $q=1,3,\dots, N_k-1$ with distribution
      \begin{equation} 
	    e^{-\frac{N_k}{4} 2 [y_{qN/N_k} -\frac{1}{2}(y_{(q+1)N/N_k}+
		y_{(q-1)N/N_k})]^2}. \label{ynk}
	  \end{equation}

      \item[(4)] Terminate the procedure if $N_k$ has reached $N_k=N$ for
        unit loops with $N$ ppl.

      \item[(5)] For an ensemble with common center of mass, shift each
        whole loop accordingly.

    \end{itemize}
\end{quote}

The above d-loop algorithm was selected since it is simple and about
10\% faster than previous algorithms, according to its developers,
because it requires fewer algebraic operations.  The generation of the
loops is largely independent from the main program.  Because of this,
it was simpler to generate the loops using a Matlab script.
This function was used to produce text files containing the worldline data for 
ensembles of loops. Then, these text files were read into memory at the 
launch of each calculation. The results of this generation routine can 
be seen in figure \ref{fig:worldlineplot}.

When the \ac{CUDA} kernel is called\footnote{Please see appendix
  \ref{ch:cudafication} for an overview of \ac{CUDA}.}, every thread
in every block executes the kernel function with its own unique
identifier. Therefore, it is best to generate worldlines in integer
multiples of the number of threads per block. The Tesla C1060 device
allows up to 512 threads per block.
\begin{figure}
	\centering
		\includegraphics[width=\linewidth,clip,trim=2cm 1cm 2cm 1cm]{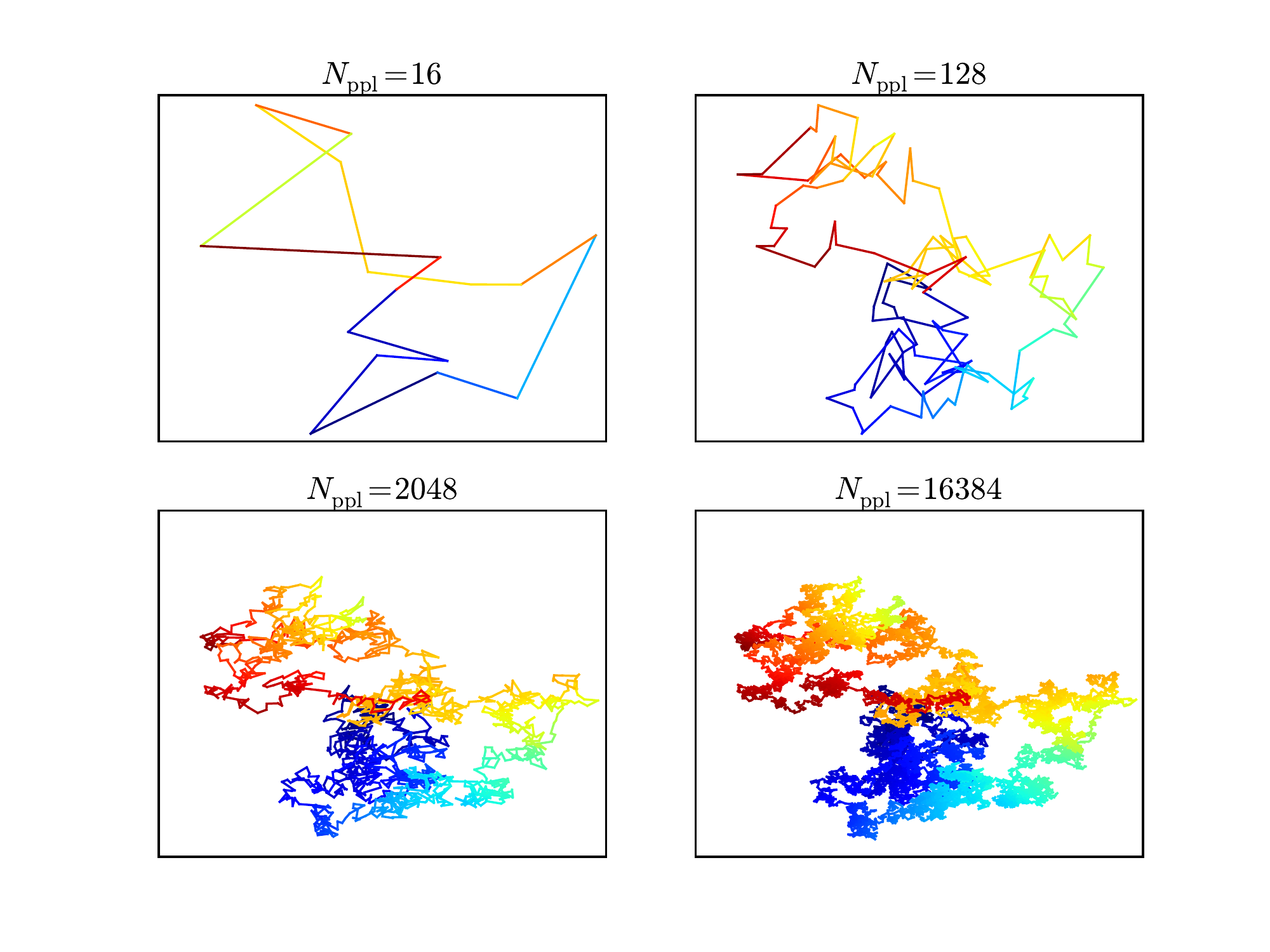}
	\caption[Discretization of the worldline loop]
	{A single discrete worldline loop shown at several levels 
	of discretization. The loops form fractal patterns and have a strong 
	parallel with Brownian motion. The colour
	represents the phase of a particle travelling along the loop, and 
	begins at dark blue, progresses in a random walk through yellow, 
	and ends at dark 
	red. The total flux through this particular worldline at $T=1$ and 
	$B=B_k$ is about $0.08 \pi/e$.}
	\label{fig:worldlineplot}
\end{figure}

\section{Cylindrical Worldline Numerics} 
We now consider cylindrically symmetric external magnetic fields.
In this case, we may simplify (\ref{eqn:QEDWL}),
\ba 
  \label{eqn:cylEA}
\frac{\Gamma^{(1)}_\mathrm{ ferm}}{T
L_z} &=& \frac{1}{4\pi} \int_0^\infty \rho_\mathrm{ cm}
  \biggr[ \int_0^\infty \frac{dT}{T^3}e^{-m^2T} \times \\
& & ~~~~~~ \left\{\langle
  W\rangle_{\vec{r}_\mathrm{ cm}} - 1 -\frac{1}{3}(eB_\mathrm{
  cm}T)^2\right\}\biggr]d\rho. \nonumber
\ea 

\subsection{Cylindrical Magnetic Fields} 

We have $\vec{B} =
B(\rho)$\bhattext{z} with
\be \label{eqn:BWLN} B(\rho) = \frac{A_\phi(\rho)}{\rho} +
\frac{dA_\phi(\rho)}{d\rho} \ee if we make the gauge choice that $A_0 =
A_\rho = A_z = 0$.

We begin by considering $\Aphi$ in the form
\be \Aphi = \frac{F}{2\pi \rho}f_\lambda(\rho) \ee so that \be
B_z(\rho)=\frac{F}{2\pi\rho}\frac{df_\lambda(\rho)}{d\rho} \ee and the total
flux is \be \Phi=F(f_\lambda(L_\rho)-f_\lambda(0)).  \ee It is convenient
to express the flux in units of $\frac{2 \pi}{e}$ and define a dimensionless
quantity
\be 
	\mathcal{F}=\frac{e}{2 \pi} F.
\ee







\subsection{Wilson Loop}

The quantity inside the angled brackets in equation (\ref{eqn:QEDWL}) is a 
gauge invariant observable called a Wilson loop. We note that the proper time
integral provides a natural path ordering for this operator.
The Wilson loop expectation value is
\be
\label{eqn:wilsonloop}
\langle W\rangle_{\vec{r}_\mathrm{cm}} \!\! = \biggl \langle e^{ie\int_0^T d\tau
\vec{A}(\vec{r}_{\mathrm{ cm}} + \vec{r}(\tau)) \cdot \dot{\vec{r}}}
 \frac{1}{4} \mathrm{ tr}  e^{\frac{e}{2}\int_0^T d\tau
  \sigma_{\mu \nu}F_{\mu \nu}(\vec{r}_{\mathrm{ cm}} + \vec{r}(\tau))}\biggr
  \rangle_{\vec{r}_\mathrm{ cm}} \!\!\!\!\!\!\!\!,
\ee
which we look at as a product between a scalar part ($e^{ie\int_0^T d\tau
\vec{A}(\vec{r}_{\mathrm{ cm}} + \vec{r}(\tau)) \cdot \dot{\vec{r}}}$)
and a fermionic part ($\frac{1}{4} \mathrm{ tr}  e^{\frac{e}{2}\int_0^T d\tau
  \sigma_{\mu \nu}F_{\mu \nu}(\vec{r}_{\mathrm{ cm}} + \vec{r}(\tau))}$).

\subsubsection{Scalar Part}

In a magnetic field, the scalar part is related to the flux through 
the loop, $\Phi_B$, by Stokes theorem:
\ba
 e^{ie\int_0^T d\tau
 \vec{A} \cdot \dot{\vec{r}}} &=&
 e^{ie\oint \vec{A}\cdot d\vec{r}} = e^{ie\int_{\vec{\Sigma}} \vec{\nabla}\times\vec{A} \cdot d\vec{\Sigma}}\\
 & = & e^{ie\int_{\vec{\Sigma}} \vec{B} d\vec{\Sigma}} = e^{ie\Phi_B}.
\ea
Consequently, this factor accounts for the Aharonov-Bohm phase acquired by 
particles in the loop.

The loop discretization results in the following approximation of the
scalar integral:
\be
  \oint \vec{A}(\vec{r})\cdot d\vec{r} =  \sum_{i=1}^{N_\mathrm{ppl}}
  \int_{\vec{r}^i}^{\vec{r}^{i+1}}\vec{A}(\vec{r})\cdot d\vec{r}.
\ee 
Using a linear parameterization of the positions, the line integrals are
\be 
  \int_{\vec{r}^i}^{\vec{r}^{i+1}}\vec{A}(\vec{r})\cdot d\vec{r} =
  \int_0^1dt \vec{A}(\vec{r}(t))\cdot(\vec{r}^{i+1} - \vec{r}^i).
\ee
Using the same gauge choice outlined above ($\vec{A}=A_\phi \hat{\phi}$),
we may write
\be 
	\vec{A}(\vec{r}(t)) = \frac{\mathcal{F}}{e\rho^2} 
	f_\lambda(\rho^2)(-y,x,0),
\ee 
where we have chosen $f_\lambda(\rho^2)$ to depend on $\rho^2$ instead
of $\rho$ to simplify some expressions and to 
avoid taking many costly square roots in the worldline numerics.
We then have
\be \int_{\vec{r}^i}^{\vec{r}^{i+1}}\vec{A}(\vec{r})\cdot
d\vec{r} = \mathcal{F} (x^iy^{i+1}-y^i x^{i+1})\int_0^1
dt\frac{f_\lambda(\rho_i^2(t))}{\rho_i^2(t)}.  \ee The linear interpolation
in Cartesian coordinates gives
\be 
	\label{eqn:rhoi} \rho_i^2(t) = A_i + 2B_it + C_i t^2,
\ee
where
\ba 
	A_i &=& (x^i)^2 + (y^i)^2 \\ 
	B_i&=&x^i(x^{i+1} - x^i) + y^i(y^{i+1}-y^i)\\ 
	C_i &= &(x^{i+1}-x^i)^2 + (y^{i+1}-y^i)^2. 
	\label{eqn:Ci} 
\ea

In performing the integrals along the straight lines connecting
each discretized loop point, we are in danger of violating gauge invariance.
If these integrals can be performed analytically, than gauge invariance
is preserved exactly.  However, in general, we wish to compute these integrals
numerically.  In this case, gauge invariance is no longer guaranteed, but
can be preserved to any numerical precision that's desired.

\subsubsection{Fermion Part} 

For fermions, the Wilson loop is modified by a factor,
\ba 
W^\mathrm{ ferm.} &=& \frac{1}{4}\mathrm{ tr}\left(e^{\frac{1}{2}
	e\int_0^T d\tau \sigma_{\mu \nu}F^{\mu \nu}}\right)\\ 
	&=& \frac{1}{4}\mathrm{ tr}\left(e^{\sigma_{x y} 
	e\int_0^T d\tau B\left(x(\tau)\right)}\right) \\	
	&=& \cosh{\left(e \int_0^T d\tau B\left(x(\tau)\right)\right)}\\
	&=& \cosh{\left(2\mathcal{F}\int_0^T
	d\tau f'_\lambda(\rho^2(\tau))\right)},
	\label{eqn:Wfermfpl}
\ea 
where we have used the
relation 
\be 
	eB = 2\mathcal{F}\frac{d f_\lambda(\rho^2)}{d \rho^2} =
	2\mathcal{F}f'_\lambda(\rho^2).  
\ee 

This factor represents an additional contribution to the 
action because of the spin interaction with the magnetic field. 
Classically, for a particle with a magnetic moment $\vec{\mu}$ 
travelling through a magnetic field in a time $T$, the 
action is modified by a term given by
\be
	\Gamma^0_\mathrm{ spin} = \int_0^T \vec{\mu} \cdot \vec{B}(\vec{x}(\tau)) d\tau.
\ee
The magnetic moment is related to the electron spin 
$\vec{\mu} = g\left(\frac{e}{2m}\right)\vec{\sigma}$, 
so we see that the integral in the above quantum fermion factor is 
very closely related to the classical action 
associated with transporting a magnetic moment through a magnetic field:
\be
	\Gamma^0_\mathrm{ spin} = g\left(\frac{e}{2m}\right) \sigma_{x y} \int_0^T B_z(x(\tau))d\tau.
\ee
Qualitatively, we could write
\be 
	W^\mathrm{ ferm} \sim \cosh{\left(\Gamma^0_\mathrm{ spin}\right)}.
\ee

As a possibly useful aside, we may want to express 
the integral in terms of $f_\lambda(\rho^2)$ instead of its derivative.
We can do this by integrating by parts:
\ba 
	\int_0^T d\tau f'_\lambda(\rho^2(\tau))&=& 
	\frac{T}{N_\mathrm{ ppl}}\sum_{i=1}^{N_\mathrm{ ppl}}\int_0^1
        dt f'_\lambda(\rho^2_i(\tau)) \\ 
         & = & 
	\frac{T}{N_\mathrm{ ppl}}\sum_{i=1}^{N_\mathrm{ ppl}}\biggr[
	\frac{f_\lambda(\rho^2_i(t))}{2(B_i+C_it)}\biggr|^{t=1}_{t=0} \nonumber \\
        & & +\frac{C_i}{2} \!\!\int_0^1\!\!\!\!
  \frac{f_\lambda(\rho^2_i(t))}{(B_i+C_i t)^2} dt\biggr] \\ 
  & = &  \frac{T}{N_\mathrm{ ppl}}\sum_{i=1}^{N_\mathrm{
  ppl}}\frac{C_i}{2}\!\!\int_0^1\!\!\!\! \frac{f_\lambda(\rho^2_i(t))}{(B_i+C_i t)^2} dt,
\ea 
with $\rho_i^2(t)$ given by equations (\ref{eqn:rhoi}) to (\ref{eqn:Ci}).
The second equality is obtained from integration-by-parts.  In the third
equality, we use the loop sum to cancel the boundary terms in pairs:
\be 
\label{eqn:Wfermfl}
W^{\mathrm{ ferm.}} = \cosh{\left(\frac{\mathcal{F}T}{N_\mathrm{
ppl}}\sum_{i=1}^{N_{\mathrm{ ppl}}}
  C_i \int_0^1 dt \frac{f_\lambda(\rho^2_i(t))}{(B_i+C_i t)^2}\right)}.
\ee
In most cases, one would use equation (\ref{eqn:Wfermfpl}) to compute the fermion factor 
of the Wilson loop. However, 
equation (\ref{eqn:Wfermfl}) may be useful in cases where $f'_\lambda(\rho^2(\tau))$
is not known or is difficult to compute. 



\subsection{Renormalization}

The field strength renormalization counter-terms result from the small $T$
behaviour of the worldline integrand.  In the limit where $T$ is very small,
the worldline loops are very localized around their center of mass.  So,
we may approximate their contribution as being that of a constant field
with value $\vec{A}(\vec{r}_{\mathrm{ cm}})$.  Specifically, we require that the field
change slowly on the length scale defined by $\sqrt{T}$. This condition on 
$T$ can be written
\be 
T \ll \left|\frac{m^2}{e B'(\rho^2)}\right| 
= \left| \frac{m^2}{2\mathcal{F}f''_\lambda(\rho^2_\mathrm{ cm})}\right|.
\ee 

When this limit is satisfied, we may use the exact expressions for the
constant field Wilson loops to determine the small $T$ behaviour of the
integrands and the corresponding counter terms.

The Wilson loop averages for constant magnetic fields in scalar and 
fermionic \ac{QED} are
\be
	\mean{W}_\mathrm{ ferm} = eBT\coth{(eBT)}
\ee
and
\be
	\mean{W}_\mathrm{ scal} = \frac{eBT}{\sinh{(eBT)}}.
\ee
\begin{widetext}
Therefore, the integrand for fermionic \ac{QED} in the limit of small $T$ is
\ba \label{eqn:fermI} I_\mathrm{ ferm}(T) &=&
\frac{e^{-m^2T}}{T^3}\left[eB(\vec{r}_{cm})T\coth{(eB(\vec{r}_{cm})T)}
- 1 -\frac{e^2}{3} B^2(\vec{r}_{cm})T^2\right] \nonumber \\ &\approx&-\frac{(eB)^4
T}{45}+\frac{1}{45} (eB)^4 m^2 T^2+\left( \frac{2 (eB)^6}{945}-\frac{(eB)^4
m^4}{90}\right)T^3 
   +\frac{(7 (eB)^4 m^6-4 (eB)^6 m^2)T^4 }{1890}+O(T^5).
\ea
For scalar QED we have
\ba 
\label{eqn:scalI} I_\mathrm{ scal}(T) &=&
\frac{e^{-m^2T}}{T^3}\left[\frac{eB(\vec{r}_{cm})T}{\sinh{(eB(\vec{r}_{cm})T)}}
- 1 +\frac{1}{6} (eB)^2(\vec{r}_{cm})T^2\right] \nonumber \\ &\approx&\frac{7
(eB)^4 T}{360}-\frac{7  (eB)^4 m^2 T^2}{360}+\frac{(147 (eB)^4 m^4-31
(eB)^6)T^3 }{15120} +\frac{ (31 (eB)^6 m^2-49 (eB)^4
m^6)T^4}{15120}+O(T^5). 
\ea
\end{widetext}

Beyond providing the renormalization conditions, these expansions can
be used in the small $T$ regime to avoid a problem with the Wilson
loop uncertainties in this region.  Consider the uncertainty in the
integrand arising from the uncertainty in the Wilson loop:
\be
\delta I(T) = \frac{\partial I}{\partial W} \delta W = \frac{e^{-m^2 T}}{T^3} \delta W.
\ee 
In this case, even though we can compute the Wilson loops for small $T$
precisely, even a small uncertainty is magnified by a divergent factor when
computing the integrand for small values of $T$.  So, in order to perform
the integral, we must replace the small $T$ behaviour of the integrand with
the above expansions (\ref{eqn:fermI}) and (\ref{eqn:scalI}).  Our worldline
integral then proceeds by analytically computing the integral for the small
$T$ expansion up to some small value, $a$, and adding this to the remaining
part of the integral~\cite{MoyaertsLaurent:2004}:
\be 
	\int_0^{\infty} I(T) dT = \underbrace{\int_0^a I(T) dT}_\mathrm{ small ~T} 
	+ \underbrace{\int_a^\infty I(T)dT}_\mathrm{ worldline ~numerics}.
\ee
Because this normalization procedure uses the constant field expressions for small values of 
$T$, this scheme introduces a small systematic uncertainty. To improve on the 
method outlined here, the derivatives of the background field can 
be accounted for by using the analytic forms of the heat kernel expansion to perform the 
renormalization~\cite{Gies:2001tj}. 

\section{Uncertainty analysis in worldline numerics} 
\label{ch:WLError}

So far in the worldline numerics literature, the discussions of uncertainty
analysis have been unfortunately brief.  It has been suggested that the
standard deviation of the worldlines provides a good measure of the
statistical error in the worldline method~\cite{Gies:2001zp,
  Gies:2001tj}.  However, the distributions produced by the worldline
ensemble are highly non-Gaussian (see figure \ref{fig:hists}), 
and therefore the standard error in
the mean is not a good measure of the uncertainties
involved. Furthermore, the use of the same worldline ensemble to
compute the Wilson loop multiple times in an integral results in
strongly correlated uncertainties. Thus, propagating uncertainties
through integrals can be computationally expensive due to the
complexity of computing correlation coefficients.

The error bars on worldline calculations impact the conclusions that
can be drawn from calculations, and also have important implications
for the fermion problem, which limits the domain of applicability of
the technique (see section \ref{sec:fermionproblem}).  It is therefore
important that the error analysis is done thoughtfully and
transparently.  The purpose of this chapter is to contribute a more
thorough discussion of uncertainty analysis in the worldline numerics technique
to the literature in hopes of avoiding any confusion associated with
the above-mentioned subtleties.

There are two sources of uncertainty in the worldline technique: the
discretization error in treating each continuous worldline as a set of
discrete points, and the statistical error of sampling a finite number
of possible worldlines from a distribution.  In this section, we
discuss each of these sources of uncertainty.

\subsection{Estimating the Discretization Uncertainties}
\label{sec:discunc}

The discretization error arising from the integral over $\tau$ in the
exponent of each Wilson loop (see equation (\ref{eqn:wilsonloop})) is
difficult to estimate since any number of loops could be represented
by each choice of discrete points.  The general strategy is to make
this estimation by computing the Wilson loop using several different
numbers of points per worldline and observing the convergence
properties.

The specific procedure adopted for this work involves dividing each discrete worldline into several 
worldlines with varying levels of discretization.
Since we are using the d-loop 
method for generating the worldlines (section \ref{sec:loopgen}), 
a $\frac{N_\mathrm{ppl}}{2}$ sub-loop consisting of every other 
point will be guaranteed to contain the prescribed distribution of velocities.

\begin{figure}
	\centering
		\includegraphics[width=\linewidth]{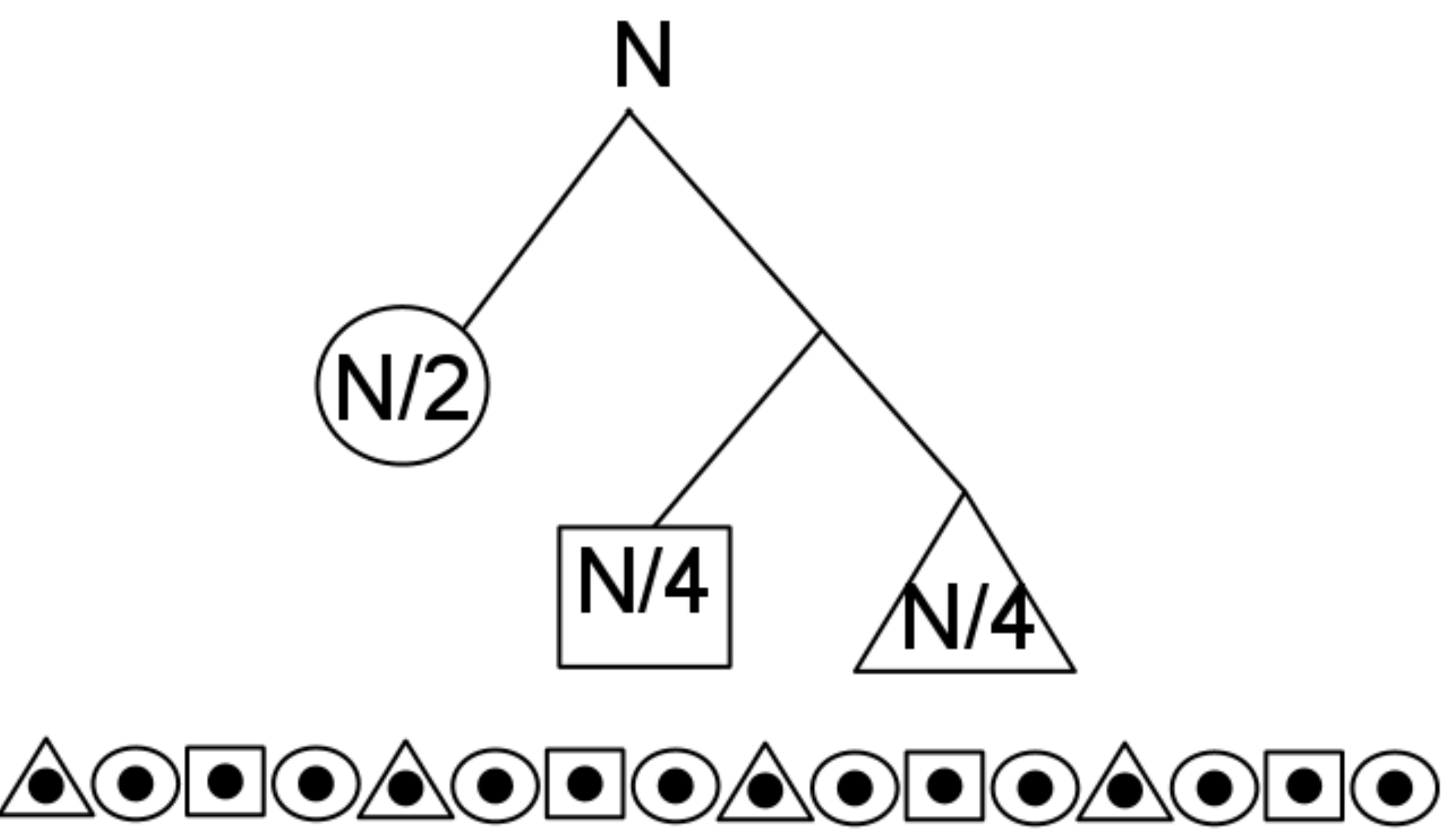}
	\caption[Illustration of convergence testing scheme]
	{Diagram illustrating the division of a worldline into three smaller interleaved worldlines}
	\label{fig:Division}
\end{figure}

To look at the convergence for the loop discretization, 
each worldline is divided into three groups.  One group of $\frac{N_\mathrm{ppl}}{2}$ points, and two groups of 
$\frac{N_\mathrm{ppl}}{4}$.  
This permits us to compute the average holonomy factors at three levels of 
discretization:
\be 
\mean{W}_{N_\mathrm{ppl}/4} = \mean{e^{\frac{i}{2}\triangle} e^{\frac{i}{2}\Box}},
\ee 
\be 
\mean{W}_{N_\mathrm{ppl}/2} = \mean{e^{i\circ}},
\ee 
and
\be 
\mean{W}_{N_\mathrm{ppl}} = \mean{e^{\frac{i}{2}\circ} e^{\frac{i}{4}\Box} e^{\frac{i}{4}\triangle}},
\ee 
where the symbols $\circ$, $\Box$, and $\triangle$ denote the worldline integral, 
$\int_0^T d\tau A(x_{CM}+x(\tau))\cdot \dot x$, computed using the sub-worldlines 
depicted in figure \ref{fig:Division}.

\begin{figure}
	\centering
		\includegraphics[width=\linewidth,clip,trim=1.8cm 0.3cm 1.8cm 1cm]{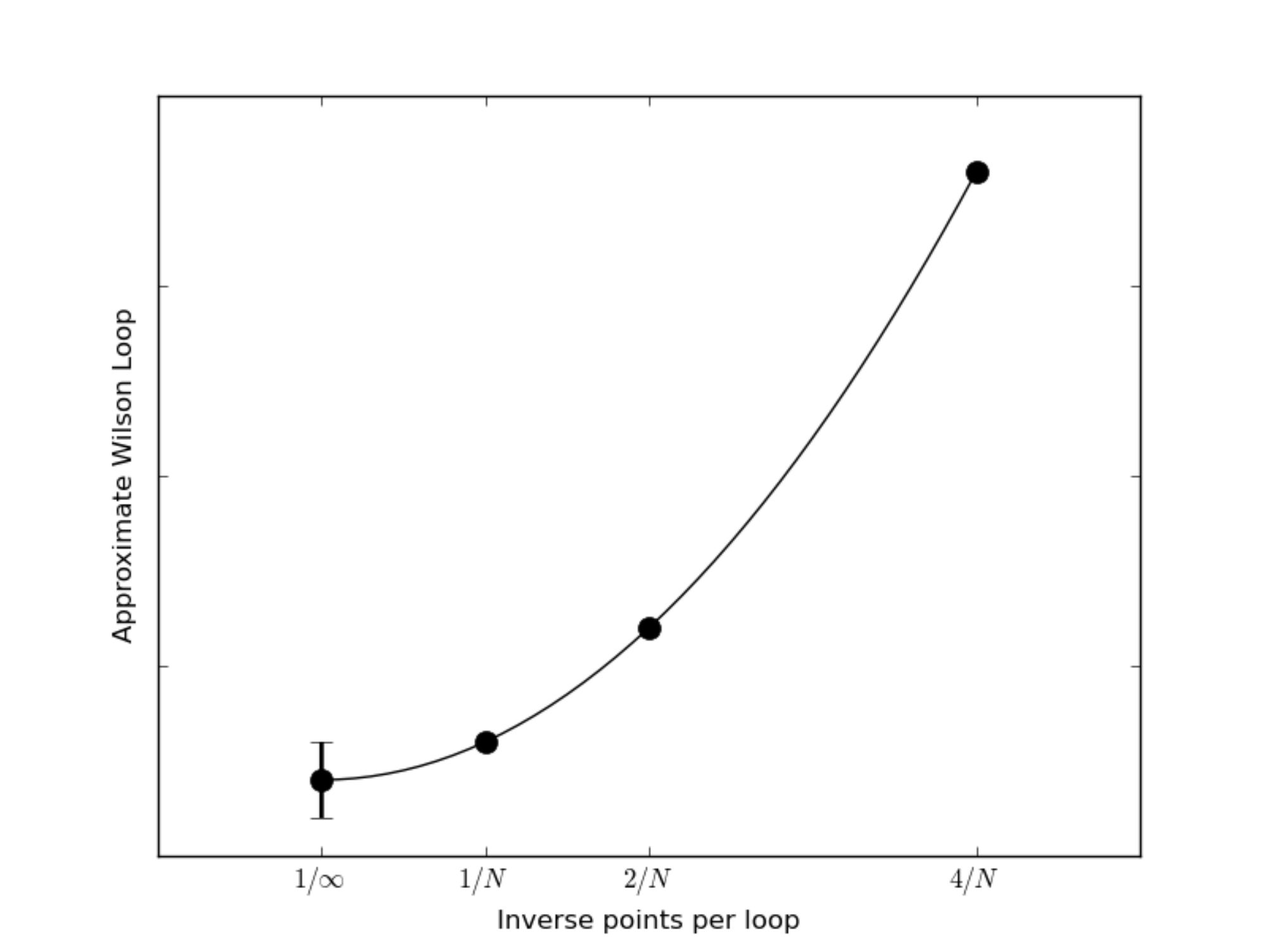}
	\caption[Illustration of extrapolation to infinite points per loop]
	{This plot illustrates the method used to extrapolate the Wilson loop to 
	infinite points per loop and the uncertainty estimate in the approximation.}
	\label{fig:DiscErr}
\end{figure}

We may put these factors into the equation of a parabola to extrapolate the result to an infinite 
number of points per line (see figure \ref{fig:DiscErr}):
\be 
\mean{W}_{\infty} \approx \frac{8}{3}\mean{W}_{N_\mathrm{ppl}} - 2 \mean{W}_{N_\mathrm{ppl}/2} +\frac{1}{3}\mean{W}_{N_\mathrm{ppl}/4}.
\ee 
So, we estimate the discretization uncertainty to be
\be 
\delta \mean{W}_{\infty} \approx |\mean{W}_{N_\mathrm{ppl}} -  \mean{W}_{\infty}|.
\ee 
Generally, the statistical uncertainties are the limitation in the precision of the 
worldline numerics technique. Therefore, $N_\mathrm{ppl}$ should be chosen to be 
large enough that the discretization uncertainties are small relative to the 
statistical uncertainties.

\subsection{Estimating the Statistical Uncertainties}

We can gain a great deal of insight into the nature of the statistical uncertainties 
by examining the specific case of the uniform magnetic field since we know the 
exact solution in this case. 

\subsubsection{The Worldline Ensemble Distribution is not Normal.}

\begin{figure}
	\centering
		\includegraphics[width=\linewidth,clip,trim=0.8cm 0.3cm 1.8cm 1cm]{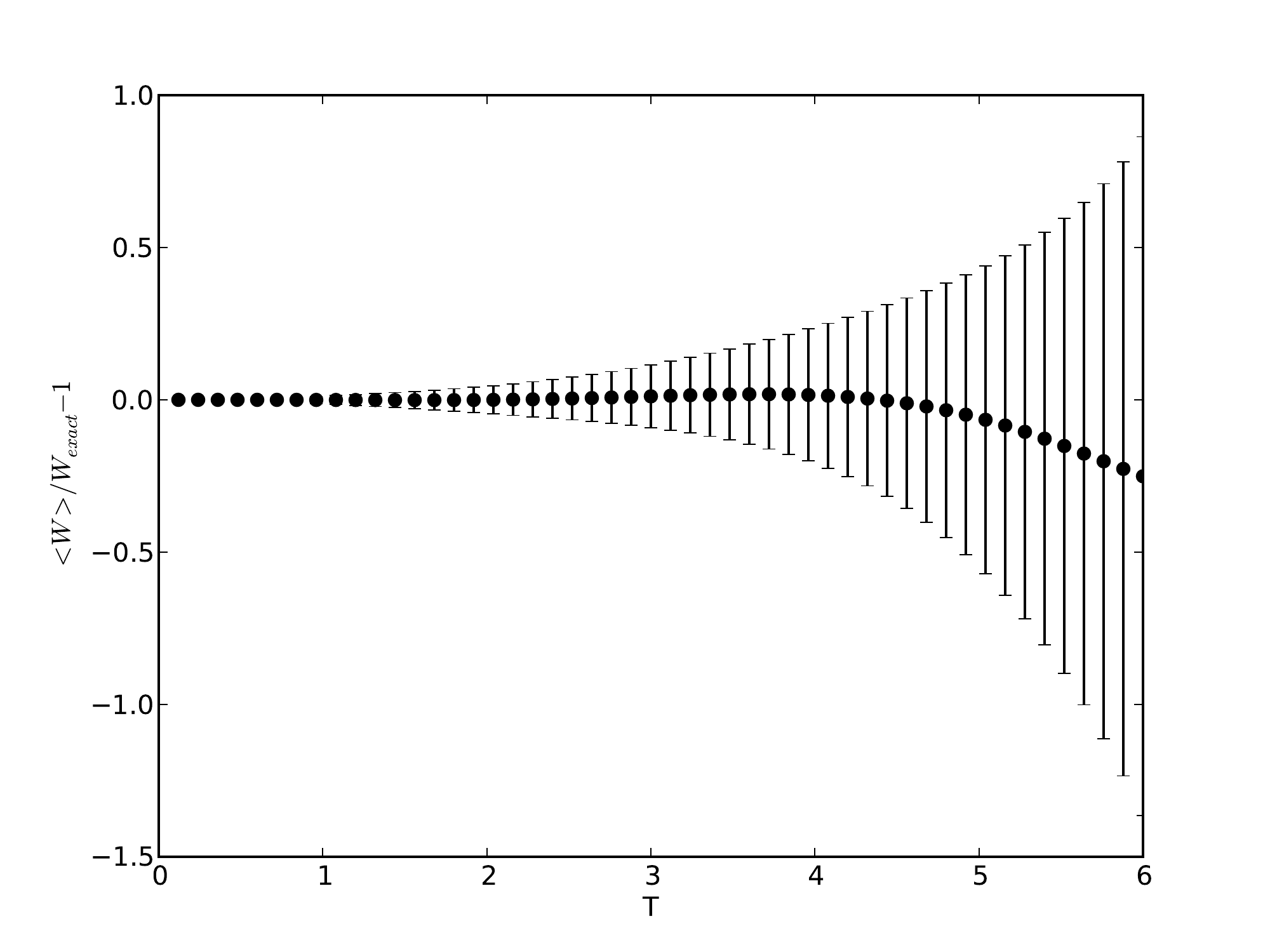}
	\caption[Standard errors in the mean for worldline numerics]
	{Residuals of worldline calculations and the corresponding standard 
	errors in the mean.  For reasons discussed in this section, these error bars 
	overestimate the uncertainties involved.}
	\label{fig:resids}
\end{figure}

A reasonable first instinct for estimating the error bars is to use the standard 
error in the mean of the collection of individual worldlines:
\be 
\mathrm{ SEM}( W ) = \sqrt{\sum_{i=1}^{N_l}\frac{(W_i - \mean{W})^2}{N_l(N_l-1)}}.
\ee 
This approach has been promoted in early papers on worldline numerics~\cite{Gies:2001zp, Gies:2001tj}.
In figure \ref{fig:resids}, we have plotted the residuals and the corresponding 
error bars for several values of the proper time parameter, $T$.  From this plot, 
it appears that the error bars are quite large in the sense that we appear to 
produce residuals which are considerably smaller than would be implied by the 
sizes of the error bars.  This suggests that we have overestimated the size of 
the uncertainty.

\begin{figure}
	\centering
		\includegraphics[width=\linewidth,clip,trim=0.8cm 0.3cm 1.8cm 1cm]{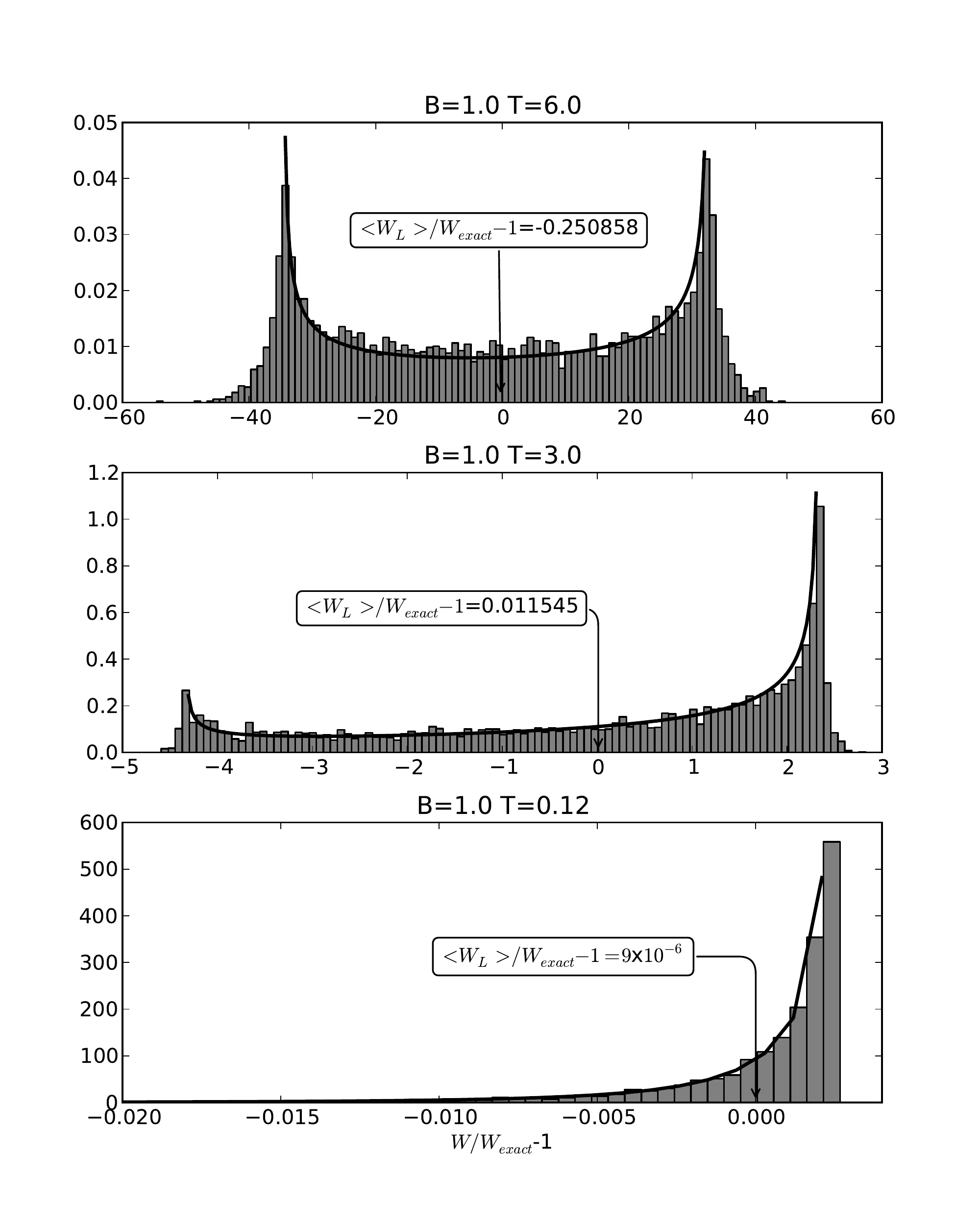}
	\caption[Histograms showing the worldline distributions]
	{Histograms showing the worldline distributions of the residuals 
	for three values of $T$ in the constant magnetic field case. 
	Here, we are neglecting the fermion factor. The dark 
	line represents the exact distribution computed using equation 
	\ref{eqn:exactdist}.  The worldline means are indicated with an arrow, 
	while the exact mean in each case is 0.  There are 5120 worldlines in 
	each histogram.  The vertical axes are normalized to a total area of unity.}
	\label{fig:hists}
\end{figure}

We can see why this is the case by looking more closely at the
distributions produced by the worldline technique.  An exact
expression for these distributions can be derived in the case of the
constant magnetic field~\cite{MoyaertsLaurent:2004}: 
\ba
\label{eqn:exactdist}
w(y)&=&\frac{W_\mathrm{ exact}}{\sqrt{1-y^2}}\sum_{n=-\infty}^{\infty}\biggl[f(\arccos(y)+2n\pi)+\nonumber \\
& & ~~~~~~~~~~~~ f(-\arccos(y)+2n\pi)\biggr]
\ea
with
\be 
f(\phi)=\frac{\pi}{4BT\cosh^2(\frac{\pi \phi}{2BT})}.
\ee 
Figure \ref{fig:hists} shows histograms of the worldline results along with the 
expected distributions.  These distributions highlight a significant hurdle in 
assigning error bars to the results of worldline numerics.

Due to their highly non-Gaussian nature, the standard error in the mean is not a good 
characterization of the distributions that are produced.  We should not interpret each individual 
worldline as a measurement of the mean value of these distributions; for large values of $BT$, 
almost all of our worldlines will produce answers which are far away from the mean of the 
distribution. This means that the variance of the distribution will be very large, even 
though our ability to determine the mean of the distribution is relatively precise
because of the increasing symmetry about the mean as $T$ becomes large.

\subsubsection{Correlations between Wilson Loops}

\begin{figure}
	\centering
		\includegraphics[width=\linewidth,clip,trim=0.8cm 0.3cm 1.8cm 1cm]{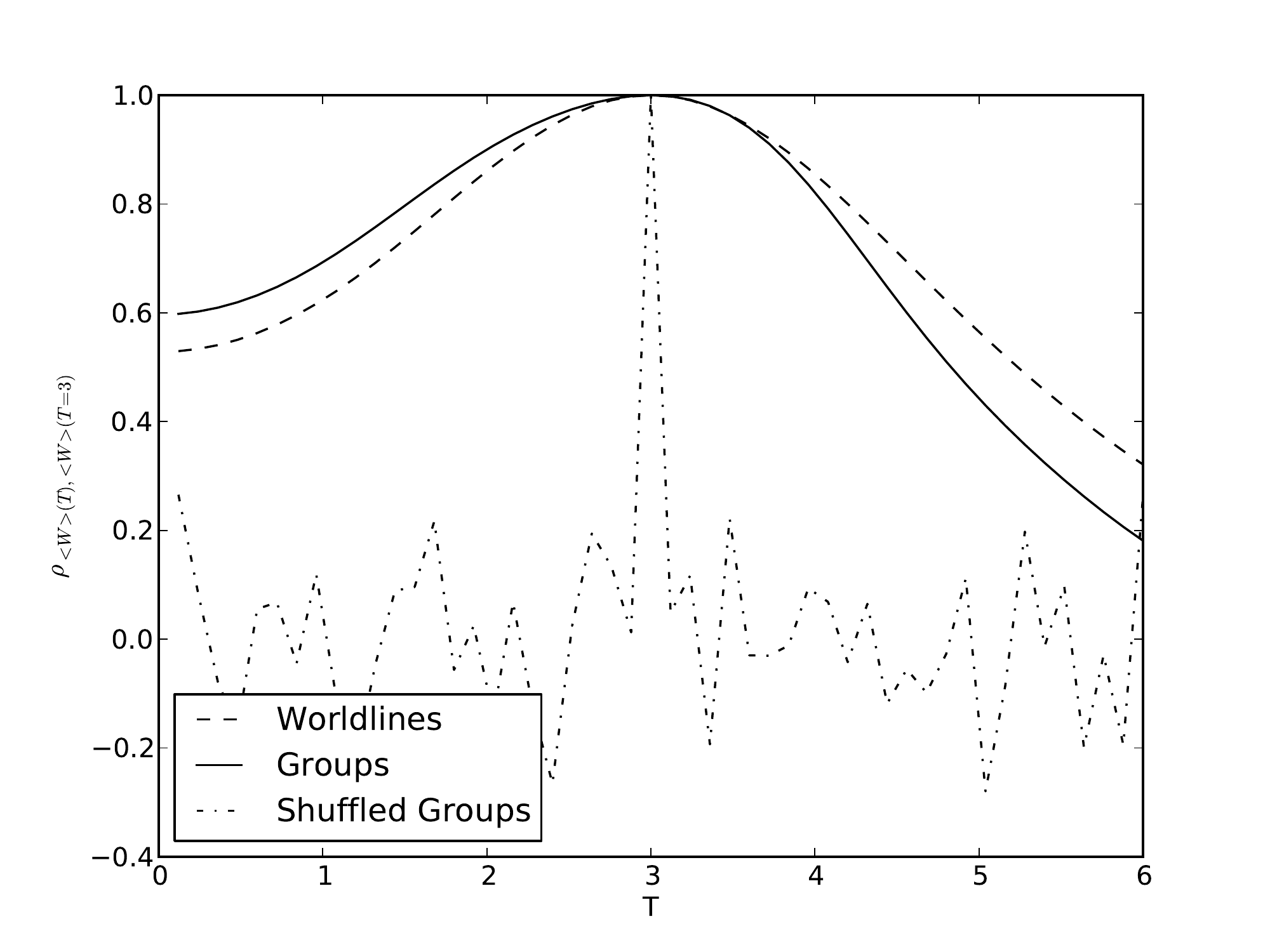}
	\caption[Correlation coefficients between different evaluations of the integrand]
	{Correlation coefficients, equation (\ref{eqn:corrcoef}), between $\mean{W(T)}$ 
		and $\mean{W(T=3)}$ computed using individual worldlines, groups of worldlines, and 
		shuffled groups of worldlines.}
	\label{fig:corr}
\end{figure}

Typically, numerical integration is performed by replacing the integral with a sum over a finite set 
of points from the integrand.  We will begin the present discussion by considering the uncertainty 
in adding together two points (labelled $i$ and $j$) in our integral over $T$. Two terms of the 
sum representing the numerical integral will involve a function of $T$ times the two 
Wilson loop factors,
\be 
I = g(T_i)\mean{W(T_i)} + g(T_j)\mean{W(T_j)}
\ee 
with an uncertainty given by
\ba 
	\delta I &=& \left | \pderiv{I}{\mean{W(T_i)}} \right|^2 (\delta \mean{W(T_i)})^2   \\ \nonumber
	& &
	~~ + \left | \pderiv{I}{\mean{W(T_j)}} \right|^2 (\delta
        \mean{W(T_j)})^2  \\ \nonumber
	& & ~~ + 2 \left | \pderiv{I}{\mean{W(T_i)}}
          \pderiv{I}{\mean{W(T_j)}} \right | \times \\ \nonumber 
& & ~~\rho_{ij} 
	(\delta \mean{W(T_i)}) (\delta \mean{W(T_j)}) \\
	&=& g(T_i)^2 (\delta \mean{W(T_i)})^2 + g(T_j)^2 (\delta \mean{W(T_j)})^2+ \nonumber \\
	& & ~ 2 \left | g(T_i)g(T_j) \right | \rho_{ij} (\delta \mean{W(T_i)}) (\delta \mean{W(T_j)})
\ea 
and the correlation coefficient $\rho_{ij}$ given by
\be 
	\label{eqn:corrcoef}
	\rho_{ij} = \frac{\mean{(W(T_i) - \mean{W(T_i)})(W(T_j)-\mean{W(T_j)})}}{\sqrt{(W(T_i)
	-\mean{W(T_i)})^2}\sqrt{(W(T_j)-\mean{W(T_j)})^2}}.
\ee 
The final term in the error propagation equation takes into account correlations between the 
random variables $W(T_i)$ and $W(T_j)$. Often in a Monte Carlo computation, one can 
treat each evaluation of the integrand as independent, and neglect the uncertainty
term involving the correlation coefficient. However, in worldline numerics, 
the evaluations are related because the same worldline ensemble is reused 
for each evaluation of the integrand.
The correlations are significant (see figure \ref{fig:corr}), and this term 
can't be neglected. Computing each correlation coefficient takes 
a time proportional to the square of the number of worldlines. Therefore, it may 
be computationally expensive to formally propagate uncertainties through an 
integral.

The point-to-point correlations were originally pointed out by Gies
and Langfeld who addressed the problem by updating (but not replacing
or regenerating) the loop ensemble in between each evaluation of the
Wilson loop average~\cite{Gies:2001zp}.  This may be a good way of
addressing the problem. However, in the following section, we promote a
method which can bypass the difficulties presented by the correlations
by treating the worldlines as a collection of worldline groups.

\subsubsection{Grouping Worldlines}

Both of the problems explained in the previous two subsections can be overcome 
by creating groups of worldline loops within the ensemble. Each group of worldlines 
then makes a statistically independent measurement of the Wilson loop average 
for that group. The statistics between the groups of measurements are normally
distributed, and so the uncertainty is the standard error in the mean of the 
ensemble of groups (in contrast to the ensemble of worldlines).

For example, if we divide the $N_l$ worldlines into $N_G$ groups of $N_l/N_G$ 
worldlines each, we can compute a mean for each group:
\be
	\mean{W}_{G_j} = \frac{N_G}{N_l}\sum_{i=1}^{N_l/N_G}W_i.
\ee
Provided each group contains the same number of worldlines, 
the average of the Wilson loop is unaffected by this grouping:
\ba
	\mean{W} & = & \frac{1}{N_G} \sum_{j=1}^{N_G} \mean{W}_{G_j} \\
		& = & \frac{1}{N_l} \sum_{i=1}^{N_l} W_i.
\ea
However, the uncertainty is the standard error in the mean of 
the groups,
\be
	\delta \mean{W} = \sqrt{\sum_{i=1}^{N_G} 
		\frac{(\mean{W}_{G_i} - \mean{W})^2}{N_G(N_G-1)}}.
\ee

Because the worldlines are unrelated to one another, the choice of how to 
group them to compute a particular Wilson loop average is arbitrary. For example, 
the simplest choice is to group the loops by the order they were generated, so that 
a particular group number, $i$, contains worldlines $iN_l/N_G$ through $(i+1)N_l/N_G -1$. 
Of course, if the same worldline groupings are used to compute different Wilson 
loop averages, they will still be correlated. We will discuss this problem in a moment.

\begin{figure}
	\centering
		\includegraphics[width=\linewidth,clip,trim=0.8cm 0.3cm 1.8cm 1cm]{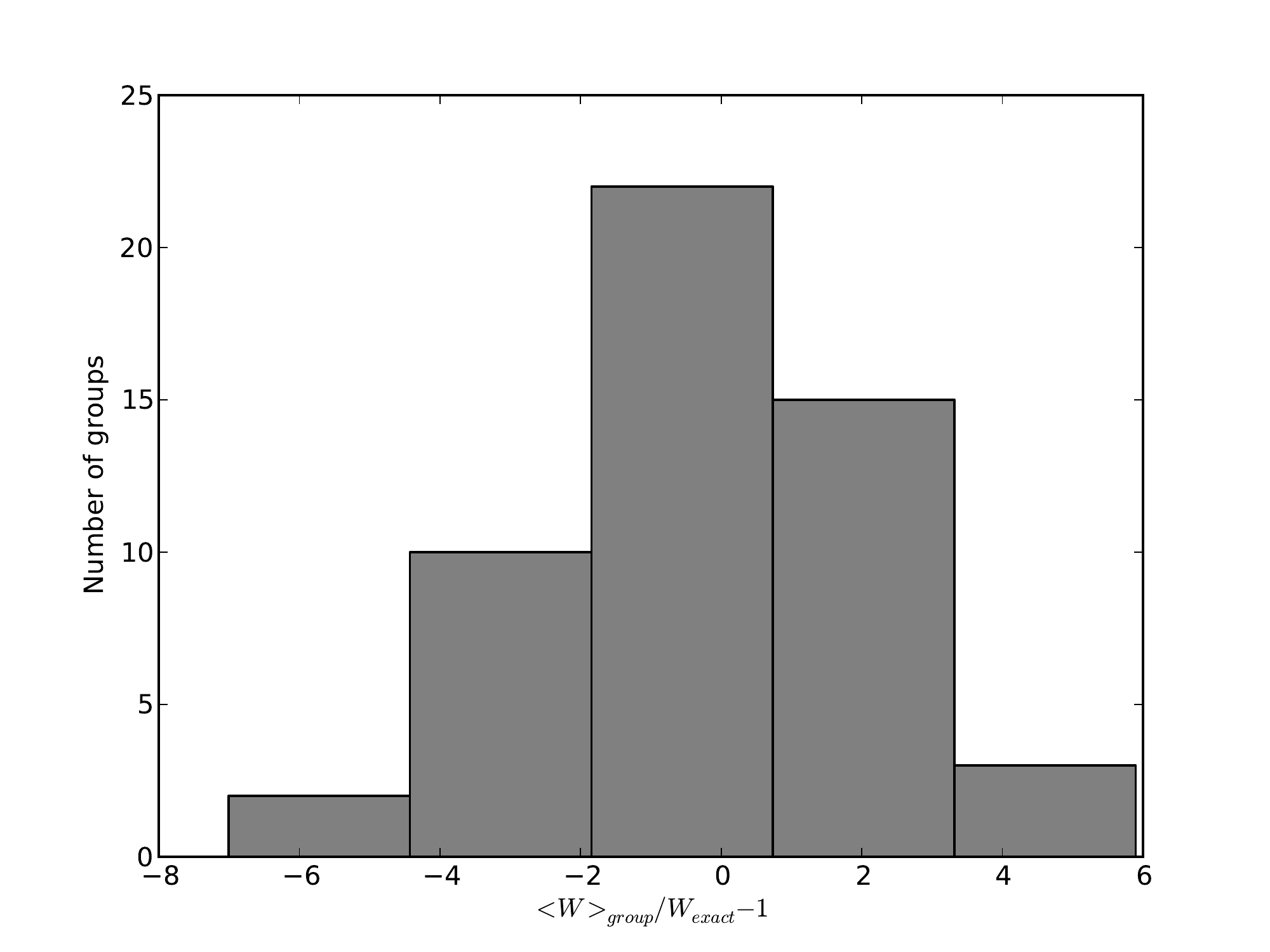}
	\caption[Histogram for reproducing measurements with groups of worldlines]
	{The histogram demonstrating the precision with which we can reproduce 
		measurements of the mean using different groups of 100 worldlines at $BT=6.0$.  
		In this case, the distribution is Gaussian-like and meaningful error bars can 
		be placed on our measurement of the mean.}
	\label{fig:uncinmean}
\end{figure}

The basic claim of the worldline technique is that the mean of the
worldline distribution approximates the holonomy factor. However, from
the distributions in figure \ref{fig:hists}, we can see that the
individual worldlines themselves do not approximate the holonomy
factor. So, we should not think of an individual worldline as an
estimator of the mean of the distribution.  Thus, a resampling
technique is required to determine the precision of our statistics. We
can think of each group of worldlines as making an independent
measurement of the mean of a distribution. As expected, the groups of
worldlines produce a more Gaussian-like distribution (see figure
\ref{fig:uncinmean}), and so the standard error of the groups is a
sensible measure of the uncertainty in the Wilson loop value.

\begin{figure}
	\centering
		\includegraphics[width=\linewidth,clip,trim=0.8cm 0.3cm 1.8cm 1cm]{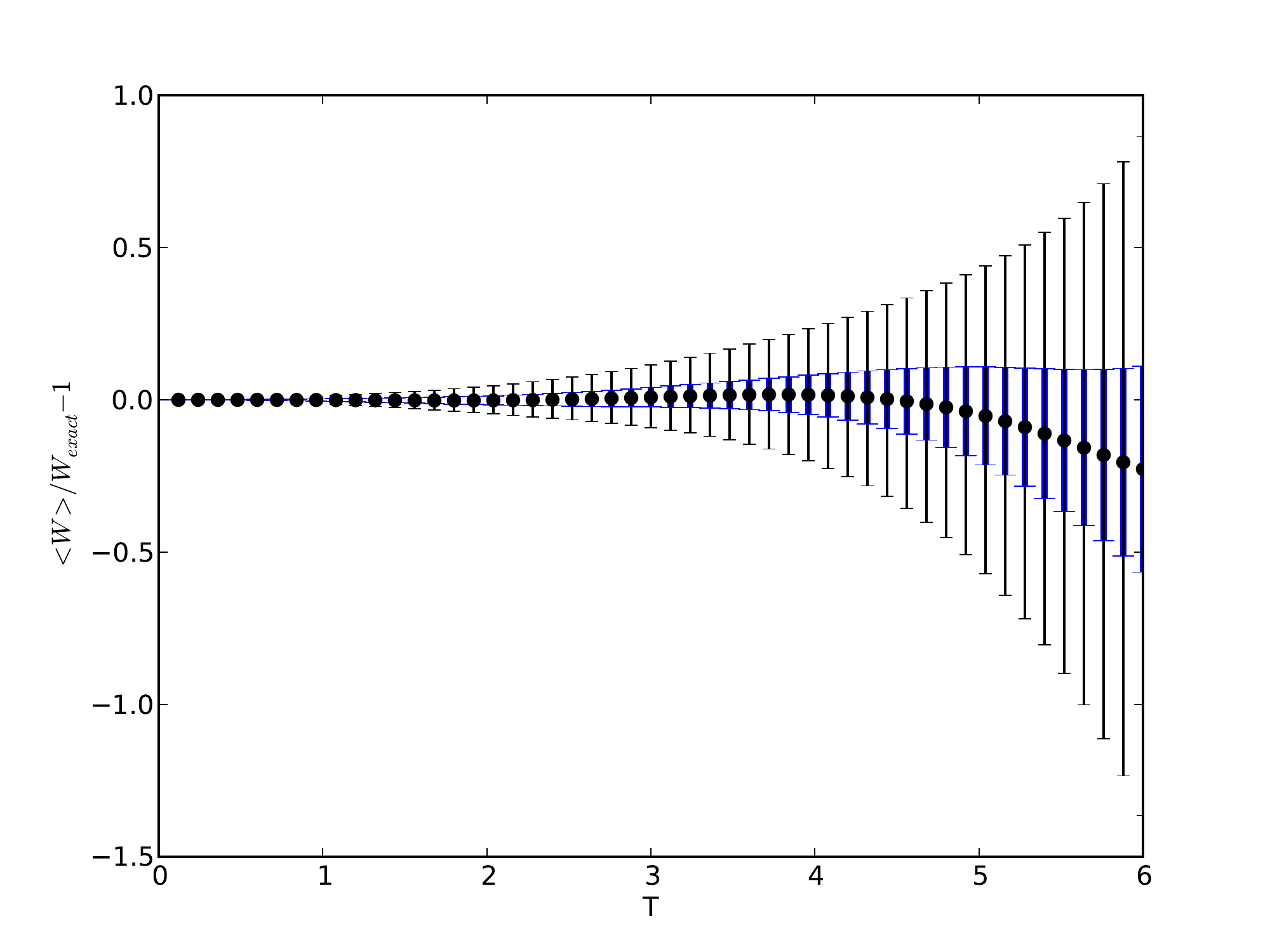}
	\caption[Comparison of error bars between standard error in the mean and 
	jackknife analysis]
	{The residuals of the Wilson loops for a constant magnetic field showing 
		the standard error in the mean (thin error bars) and the uncertainty in determining 
		the mean (thick blue error bars).  The standard error in the mean overestimates the 
		uncertainty by more than a factor of 3 at each value of $T$.}
	\label{fig:resids2}
\end{figure}

We find that the error bars are about one-third as large as those
determined from the standard error in the mean of the individual
worldlines, and the smaller error bars better characterize the size of
the residuals in the constant field case (see figure
\ref{fig:resids2}).  The strategy of using subsets of the available
data to determine error bars is called jackknifing.  Several previous
papers on worldline numerics have mentioned using jackknife analysis to
determine the uncertainties, but without an explanation of the
motivations or the procedure employed \cite{2005PhRvD..72f5001G,
  PhysRevLett.96.220401, Dunne:2009zz, PhysRevD.84.065035}.

The grouping of worldlines alone does not address the problem of
correlations between different evaluations of the integrands.  Figure
\ref{fig:corr} shows that the uncertainties for groups of worldlines
are also correlated between different points of the integrand.
However, the worldline grouping does provide a tool for bypassing the
problem. One possible strategy is to randomize how worldlines are
assigned to groups between each evaluation of the integrand. This
produces a considerable reduction in the correlations, as is shown in
figure \ref{fig:corr}. Then, errors can be propagated through the
integrals by neglecting the correlation terms. Another strategy is to
separately compute the integrals for each group of worldlines, and
then consider the statistics of the final product to determine the
error bars. This second strategy is the one adopted for the work
presented in this paper. Grouping in this way reduces the amount of
data which must be propagated through the integrals by a factor of the
group size compared to a delete-1 jackknife scheme, for example.  In
general, the error bars quoted in the remainder of this paper are obtained by
computing the standard error in the mean of groups of worldlines.

\subsection{Uncertainties and the Fermion Problem}
\label{sec:fermionproblem}

The fermion problem of worldline numerics is a name given to an enhancement of
the uncertainties at large $T$~\cite{Gies:2001zp,
  MoyaertsLaurent:2004}. It should not be confused with the
fermion-doubling problem associated with lattice methods. In a
constant magnetic field, the scalar portion of the calculation
produces a factor of $\frac{BT}{\sinh{(BT)}}$, while the fermion
portion of the calculation produces an additional factor
$\cosh{(BT)}$. Physically, this contribution arises as a result of the
energy required to transport the electron's magnetic moment around the
worldline loop.  At large values of $T$, we require subtle
cancellation between huge values produced by the fermion portion with
tiny values produced by the scalar portion.  However, for large $T$,
the scalar portion acquires large relative uncertainties which make
the computation of large $T$ contributions to the integral very
imprecise.

This can be easily understood by examining the worldline distributions shown in figure
\ref{fig:hists}. Recall that the scalar Wilson loop average for these histograms is given 
by the flux in the loop, $\Phi_B$:
\be
	\mean{W} = \left<\exp{\left(ie\int_0^Td\tau \vec{A}(\vec{x}_\mathrm{ cm} 
	+ \vec{x}(\tau))\cdot d\vec{x}(\tau)\right)}\right> = \left< e^{ie\Phi_B}\right>.
\ee
For constant fields, the flux through the worldline loops obeys the distribution function
\cite{MoyaertsLaurent:2004}
\be
	f(\Phi_B) = \frac{\pi}{4BT\cosh^2\left(\frac{\pi \Phi_B}{2BT}\right)}.
\ee
For small values of $T$, the worldline loops are small and the 
amount of flux through the loop is correspondingly small. Therefore, the 
flux for small loops is narrowly distributed about $\Phi_B = 0$. Since zero 
maximizes the Wilson loop ($e^{i0}=1$), 
this explains the enhancement to the right of the distribution for small values of $T$. 
As $T$ is increased, the flux through any given worldline becomes very large and the 
distribution of the flux becomes very broad. 
For very large $T$,  the width of the distribution is many 
factors of $2\pi/e$. Then, the phase ($e \Phi_B\mod{2\pi}$) is nearly 
uniformly distributed, and the 
Wilson loop distribution reproduces the Chebyshev distribution (\ie 
the distribution obtained from projecting uniformly distributed points on 
the unit circle onto the horizontal axis),
\be
	\lim_{T\to\infty}w(y) = \frac{1}{\pi\sqrt{1-y^2}}.
\ee

The mean of the Chebyshev distribution is zero due to its symmetry. 
However, this symmetry is not 
realized precisely unless we use a very large number of loops. Since the 
width of the distribution is already 100$\times$ the value of the mean at 
$T=6$, any numerical asymmetries in the distribution result in very large 
relative uncertainties of the scalar portion. Because of these uncertainties, 
the large contribution from the fermion factor are not cancelled precisely.

This problem makes it very difficult to compute 
the fermionic effective action unless the fields are well localized
\cite{MoyaertsLaurent:2004}. For example, the fermionic factor for 
non-homogeneous magnetic fields oriented along the z-direction is
\be
	\cosh{\left(e\int_0^T d\tau B(x(\tau))\right)}.
\ee
For a homogeneous field, this function grows exponentially with $T$ and 
is cancelled by the exponentially vanishing scalar Wilson loop.
For a localized field, 
the worldline loops are very large for large values of $T$, and they primarily 
explore regions far from the field. Thus, the fermionic factor grows more slowly 
in localized fields, and is more easily cancelled by the rapidly vanishing scalar part.

In this section, we have identified two important considerations in
determining the uncertainties associated with worldline numerics computations.
Firstly, the computed points within the integrals over proper time,
$T$, or center of mass, $\vec{x}_\mathrm{ cm}$, are highly correlated
because one typically reuses the same ensemble of worldlines to
compute each point.  Secondly, the statistics of the worldlines are
not normally distributed and each individual worldline in the ensemble
may produce a result which is very far from the mean value. So, in
determining the uncertainties in the worldline numerics technique, one should
consider how precisely the mean of the ensemble can be measured from
the ensemble and this is not necessarily given by the standard error
in the mean.

These issues can be addressed simultaneously using a scheme where the
worldlines from the ensemble are placed into groups, and the effective
action or the effective action density is evaluated separately for
each group. The uncertainties can then be determined by the statistics
of the groups.  This scheme is less computationally intensive than a
delete-1 jackknife approach because less data (by a factor of the
group size) needs to be propagated through the integrals. It is also
less computationally intensive than propagating the uncertainties
through the numerical integrals because it avoids the computation of
numerous correlation coefficients.
\begin{figure}[ht] 
  \begin{center}
    \includegraphics[width=\linewidth,clip,trim=0.8cm 0.3cm 1.8cm 1cm]{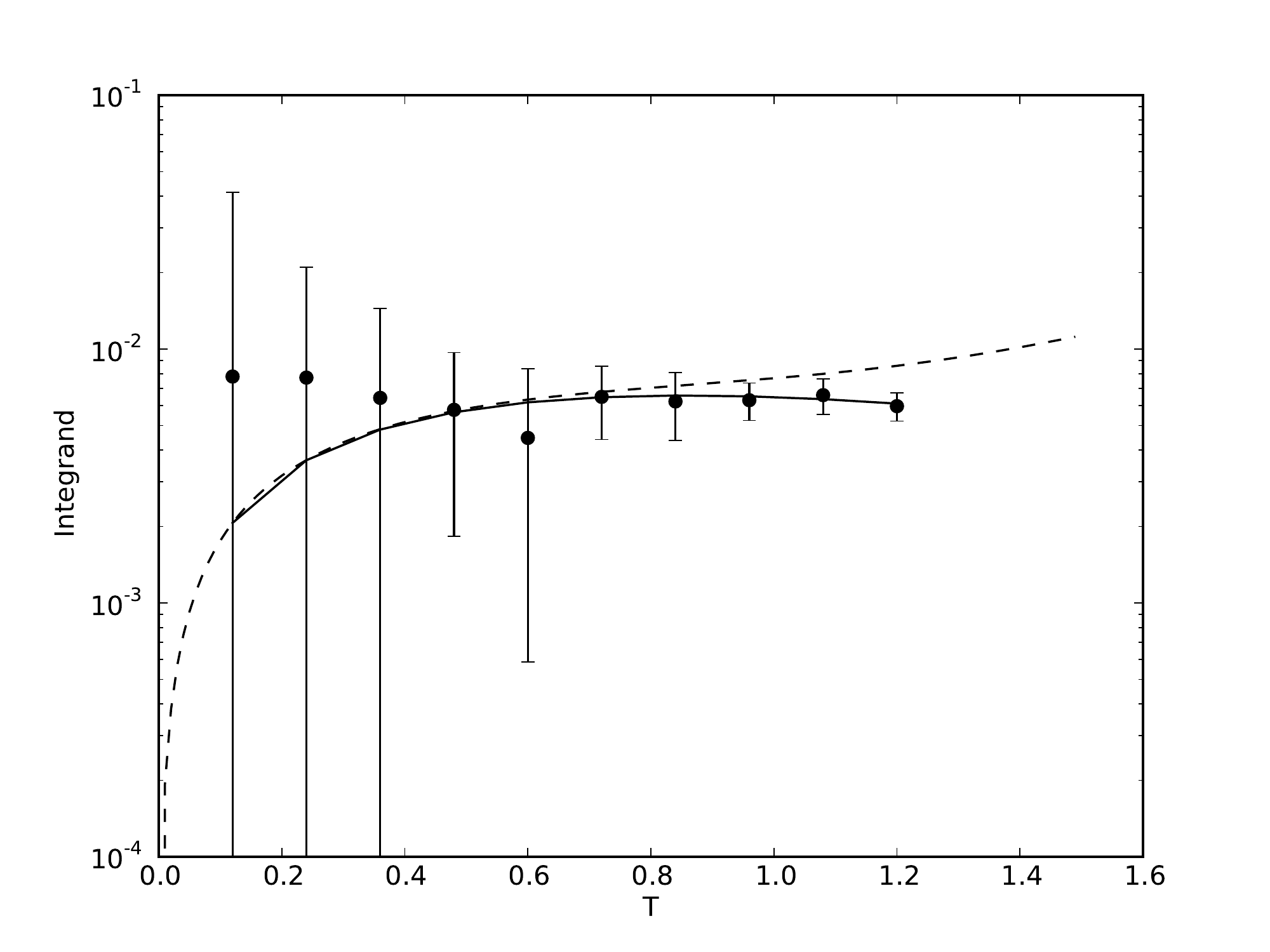} 
	    \caption[small $T$ behaviour of worldline numerics]
		{The small $T$
		behaviour of worldline numerics.  The data points represent the numerical
		results, where the error bars are determined from the jackknife analysis described 
		in chapter \ref{ch:WLError}.
		The solid line represents the exact solution while the dotted line represents
		the small $T$ expansion of the exact solution.  Note the amplification of
		the uncertainties.} 
  \end{center} 
\end{figure}

\section{Computing an Effective Action}

\begin{figure}
	\centering
		\includegraphics[width=\linewidth,clip,trim=-0.5cm 0 0 0]{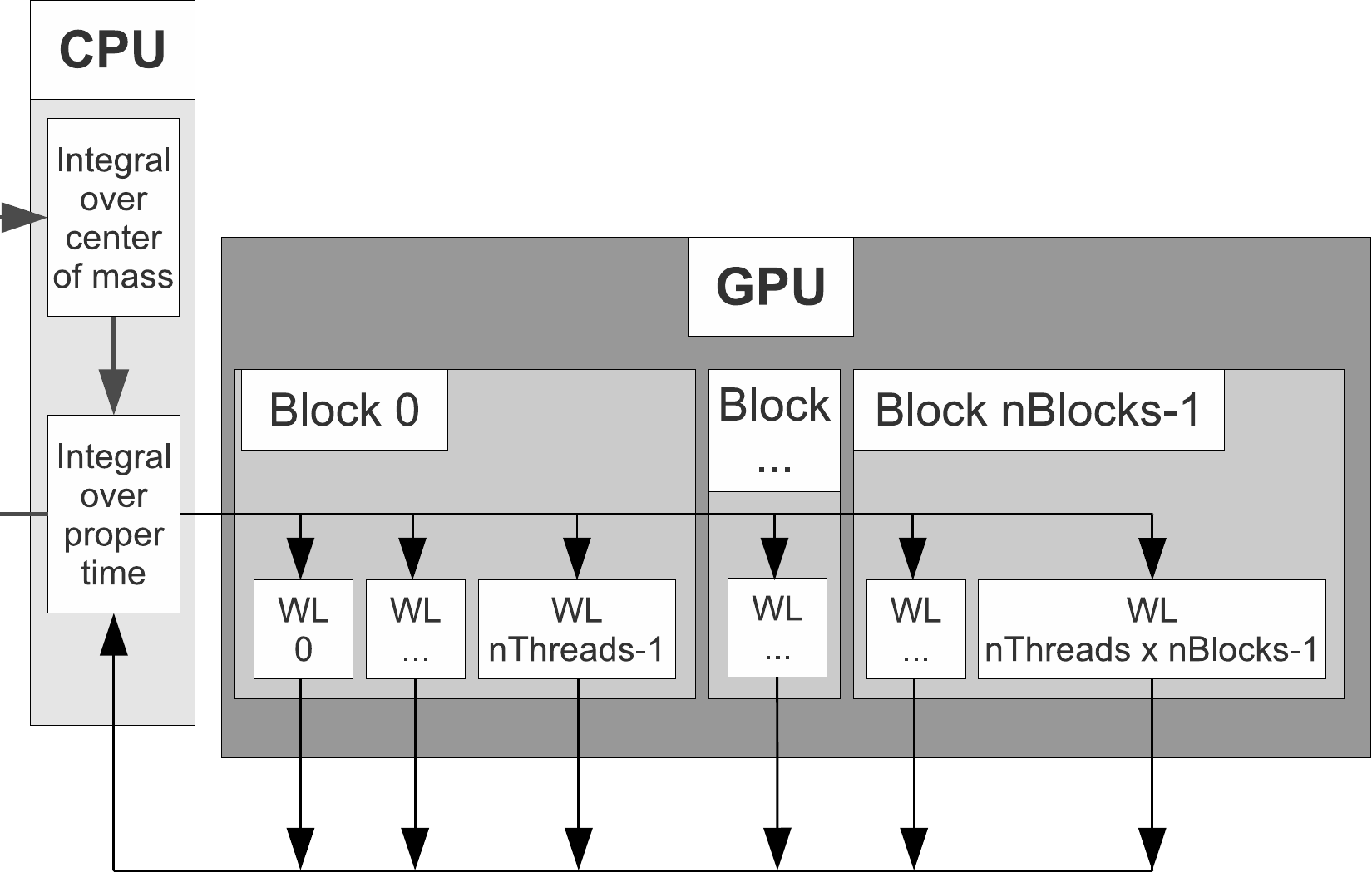}
                \caption[Heterogeneous processing scheme for worldline numerics]
				{The \ac{CPU} manages the loops which compute the 
					integrals over center of mass and proper time. 
					For each proper time integral, we require the results 
					from a large number of individual worldlines. The \ac{GPU}
					is used to compute the integral 
					over each worldline in parallel, and the 
					results are returned to the \ac{CPU} for use in the effective action
					calculation.}.
	\label{fig:coprocessing}
\end{figure}

The ensemble average in the effective action is simply the sum over the
contributions from each worldline loop, divided by the number of loops in
the ensemble.  Since the computation of each loop is independent of the
other loops, the ensemble average may be straightforwardly parallelized by
generating separate processes to compute the contribution from each loop.
For this parallelization, four Nvidia Tesla C1060 \acp{GPU} were used through
the \ac{CUDA} parallel programming framework.  Because \acp{GPU} can spawn
thousands of parallel processing threads\footnote{Each Tesla C1060 device has 30 multiprocessors
which can each support 1,024 active threads.  So, the number of threads available at a time 
is 30,720 on each of the Tesla devices.  Billions of threads may be scheduled on each 
device~\cite{cudazone}.} 
with much less computational overhead
than an \ac{MPI} cluster, they excel at handling a very large number of parallel 
threads, although the clock speed is slower and fewer memory resources are typically available.
In contrast, parallel computing on a cluster using \acs{CPU} 
tends to have a much higher speed per thread, but there are typically fewer 
threads available.  The worldline technique is exceedingly parallelizable, 
and it is a straightforward matter to divide the task into thousands or tens of thousands of 
parallel threads.  In this case,  one should expect excellent performance 
from a \ac{GPU} over a parallel \ac{CPU} implementation, unless thousands 
of \acp{CPU} are available for the program. The \ac{GPU} architecture has 
recently been used by another group for computing Casimir forces using 
worldline numerics~\cite{2011arXiv1110.5936A}.
Figure \ref{fig:coprocessing} 
illustrates the co-processing and parallelization scheme used here for the 
worldline numerics.

In an
informal test, a Wilson loop average was computed using an ensemble of 5000
loops in 4.7553 seconds using a serial implementation 
while the \ac{GPU} performed the same calculation
in 0.001315 seconds using a \ac{CUDA} implementation.  
A parallel \ac{CPU} code would require a cluster with thousands of 
cores to achieve a similar speed, even if we assume linear (ideal) 
speed-up.  So, for worldline numerics computations,
a relatively inexpensive \ac{GPU} can be expected to outperform a small
or mid-sized cluster.  This increase in computation speed has enabled the
detailed parameter searches discussed in this dissertation.

Of course, there are also trade-offs from using the 
\ac{GPU} architecture with the worldline numerics technique. The most significant 
of these is the limited availability of memory on the device. A \ac{GPU} device 
provides only a few GB of global memory (4GB on the Tesla C1060). This limit 
forces compromises between the number of points-per-loop and the number of 
loops to keep the total size of the loop cloud data small. The limited availability of memory 
resources also limits the number of threads which can be executed concurrently on the device.
Because of the overhead associated with transferring data to and from the device, 
the advantages of a \ac{GPU} over a \ac{CPU} cluster are most pronounced 
on problems which can be divided into several hundred or thousands of processes. 
Therefore, the \ac{GPU} may not offer great performance advantages 
for a small number of loops. Finally, there is some additional complexity involved 
in programming for the \ac{GPU} in terms of learning specialized libraries and 
memory management on the device. This means that the code may take longer to develop
and there may be a learning curve for researchers. However, this problem 
is not much more pronounced with \ac{GPU} programming than with other 
parallelization strategies.

Once the ensemble average of the Wilson loop has been computed,
computing the effective action is a straightforward matter of
performing numerical integrals.  The effective action density is
computed by performing the integration over proper time, $T$.  Then,
the effective action is computed by performing a spacetime integral
over the loop ensemble center of mass.  In all cases where a numerical
integral was performed, Simpson's method was
used~\cite{burden2001numerical}. Integrals from 0 to $\infty$ were
mapped to the interval $[0,1]$ using substitutions of the form $x =
\frac{1}{1+T/T_\mathrm{ max}}$, where $T_\mathrm{ max}$ sets the scale
for the peak of the integrand. In the constant field case, for the
integral over proper time, we expect $T_\mathrm{ max} \sim 3/(eB)$ for
large fields and $T_\mathrm{ max} \sim 1$ for fields of a few times
critical or smaller. In section~\ref{ch:WLError}, we presented a
detailed discussion of how the statistical and discretization
uncertainties can be computed in this technique.

In this implementation, the numerical integrals are done using a
serial CPU computation.  This serial portion of the algorithm tends to
limit the speedup that can be achieved with the large number of
parallel threads on the \ac{GPU} device\footnote{By Amdahl's law, for
  a program with a ratio, $P$, of parallel to serial computations on
  $N$ processors, the speedup is given by $S <
  \frac{1}{(1-P)+\frac{P}{N}}$. This law predicts rapidly diminishing
  returns from increasing the number of processors when $P>0$ for a
  fixed problem size.}.  However, an important benefit of the large
number of threads available on the \ac{GPU} is that the number of
worldlines in the ensemble can be increased without limit, as long as
more threads are available, without significantly increasing the
computation time.  If perfect occupancy could be achieved on a Tesla
C1060 device, an ensemble of up to 30,720 worldlines could be computed
concurrently.  Thus, the \ac{GPU} provides an excellent architecture
for improving the statistical uncertainties. Additionally, there is
room for further optimization of the algorithm by parallelizing the
serial portions of the algorithm to achieve a greater speedup.

More details about implementing this algorithm on the \ac{CUDA}
architecture can be found in appendix~\ref{ch:cudafication}. A listing
of the \ac{CUDA} worldline numerics code appears in appendix of
\cite{2012PhDT........21M}.

\section{Verification and Validation}

The worldline numerics software can be validated and verified by making sure that it 
produces the correct results where the derivative expansion is a good approximation, 
and that the results are consistent with previous numerical calculations of flux tube 
effective actions. For this reason, the validation was done primarily with flux tubes with 
a profile defined by the function
\be
	f_\lambda(\rho^2) = \frac{\rho^2}{(\lambda^2 + \rho^2)}.
\ee
For large values of $\lambda$, this function varies slowly on the Compton wavelength 
scale, and so the derivative expansion is a good approximation. Also, flux tubes 
with this profile were studied previously using worldline numerics
\cite{Moyaerts:2003ts, MoyaertsLaurent:2004}.

Among the results presented in~\cite{MoyaertsLaurent:2004} is a comparison of 
the derivative expansion and worldline numerics for this magnetic field 
configuration. The result is that the next-to-leading-order term 
in the derivative expansion is only a small correction to the the 
leading-order term for $\lambda \gg \lambda_e$, where the derivative 
expansion is a good approximation. The derivative expansion breaks
down before it reaches its formal validity limits 
at $\lambda \sim \lambda_e$. For this reason,
we will simply focus on the leading order derivative expansion (which we call 
the \ac{LCF} approximation).
The effective action of \ac{QED} in the \ac{LCF} approximation is given 
in cylindrical symmetry by
\ba
	\label{eqn:LCFferm}
	\Gamma^{(1)}_\mathrm{ ferm} &=& \frac{1}{4\pi}\int_0^\infty dT 
	\int_0^\infty \rho_\mathrm{ cm} d\rho_\mathrm{ cm}\frac{e^{-m^2T}}{T^3} \times \nonumber \\
	& &\biggl\{eB(\rho_\mathrm{ cm})T\coth{(eB(\rho_\mathrm{ cm})T)} \\
        & & 
	- 1 -\frac{1}{3}(eB(\rho_\mathrm{ cm})T)^2\biggr\}. \nonumber
\ea

\begin{figure}
	\centering
		\includegraphics[width=\linewidth,clip,trim=0.8cm 0.3cm 1.2cm 1cm]{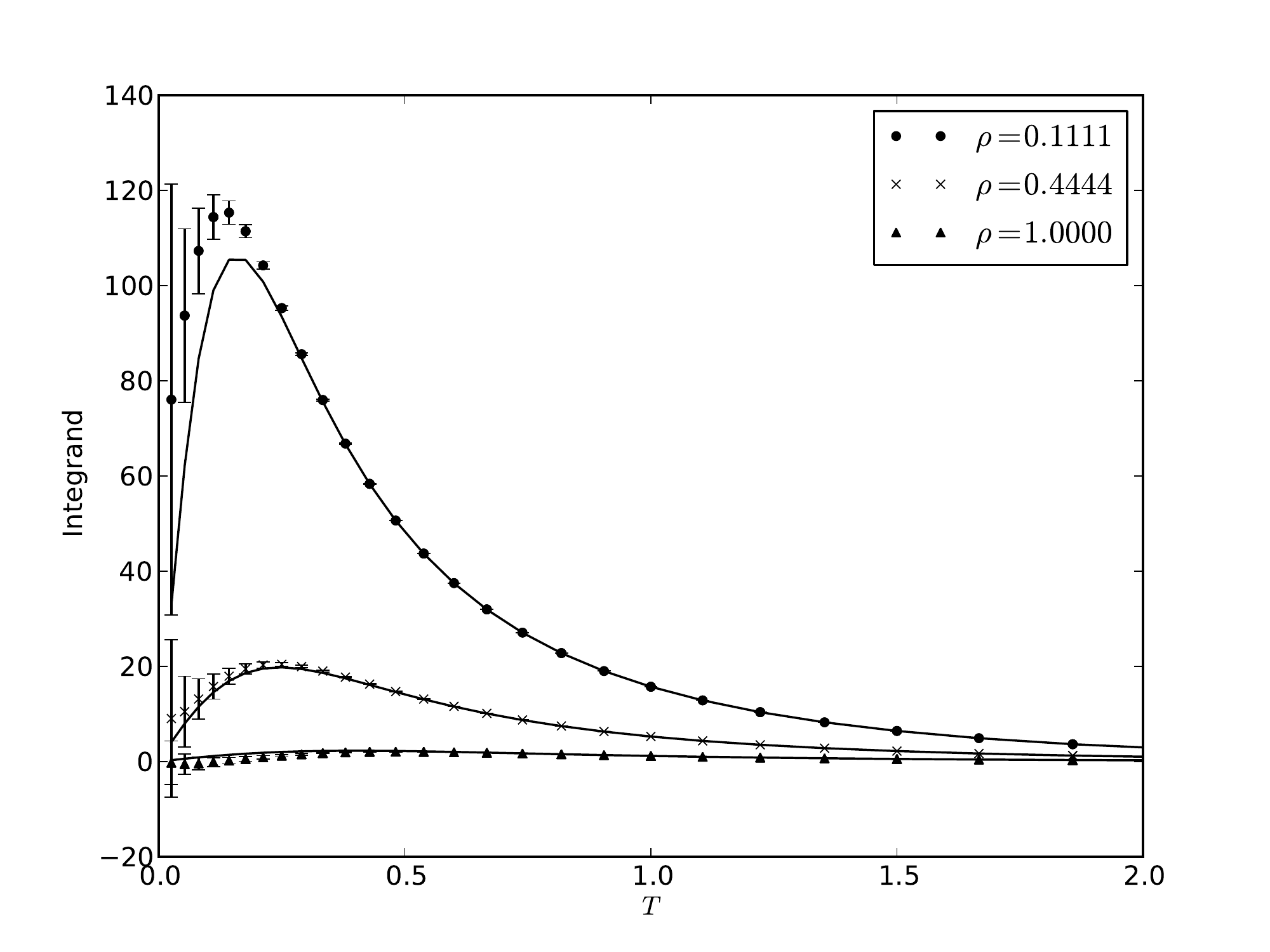}
	\caption[Comparison with derivative expansion for $T$ integrand]
	{The integrand of the proper time, $T$, integral for three different values 
	of the radial coordinate, $\rho$ for a $\lambda = 1$ flux tube. The solid lines 
	represent the zeroth-order derivative expansion, which, as expected, is a good approximation 
	until $\rho$ becomes too small.}
	\label{fig:igrandvsT}
\end{figure}

Figure \ref{fig:igrandvsT} shows a comparison between the proper time integrand,
\be
	\frac{e^{-m^2T}}{T^3}\left[\langle W\rangle_{\vec{r}_\mathrm{ cm}} 
		- 1 -\frac{1}{3}(eB_\mathrm{ cm}T)^2\right],
\ee
and the \ac{LCF} approximation result for a flux tube with $\lambda = \lambda_e$ and
$\mathcal{F} = 10$. In this case, the \ac{LCF} approximation is only appropriate far from the 
center of the flux tube, where the field is not changing very rapidly. In the figure, we can 
begin to see the deviation from this approximation, which gets more pronounced closer to the 
center of the flux tube (smaller values of $\rho$).
\begin{figure}
	\centering
		\includegraphics[width=\linewidth,clip,trim=0.2cm 0.3cm 1.8cm 1cm]{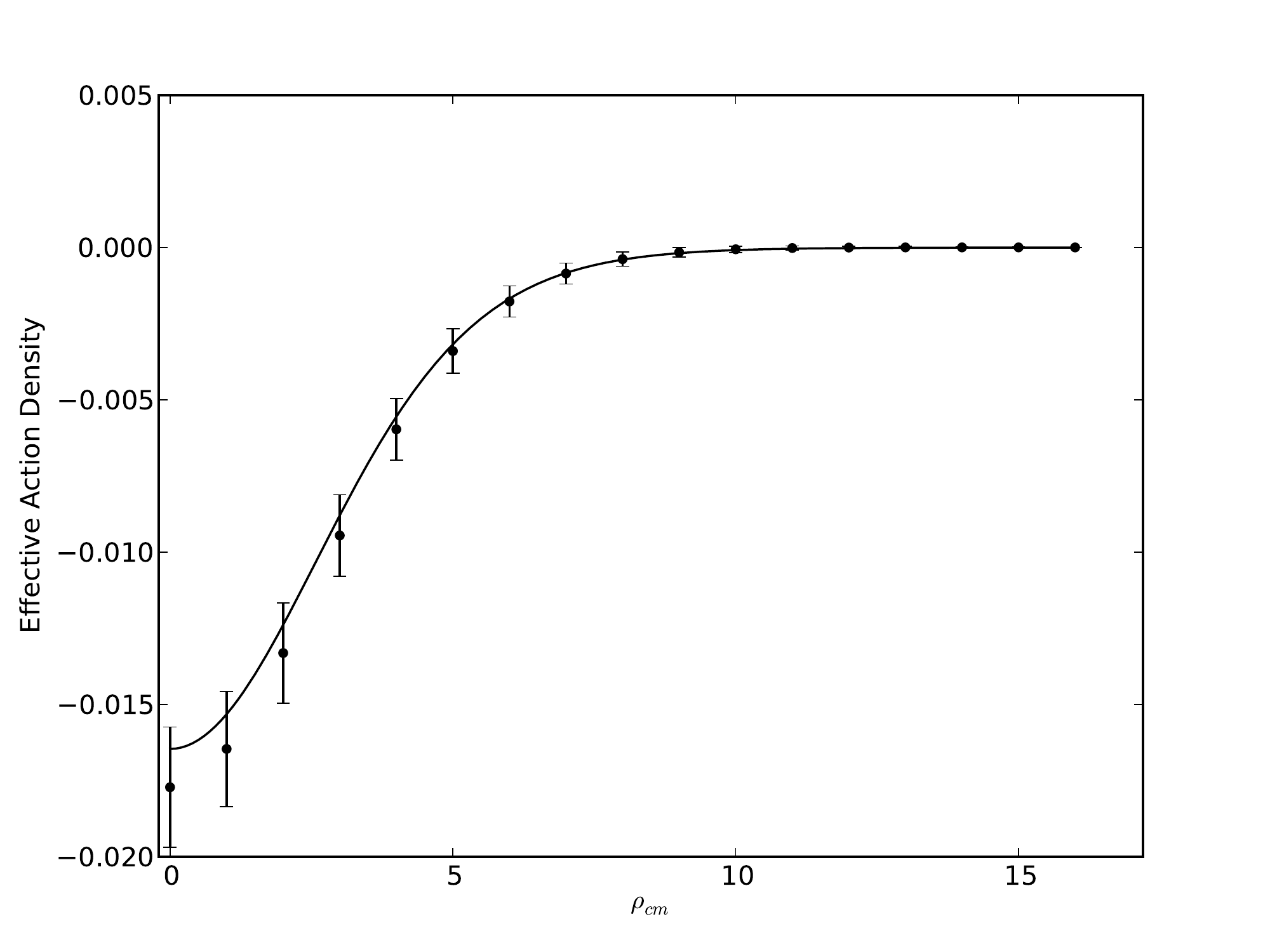}
	\caption[Comparison with \acs{LCF} approximation for action density]
	{The fermion term of the effective action density as a function 
	of the radial coordinate for a flux tube with width $\lambda = 10 \lambda_e$. }
	\label{fig:vsrhocm}
\end{figure}

The effective action density for a slowly varying flux tube is plotted in figure
\ref{fig:vsrhocm} along with the \ac{LCF} approximation. In this case, 
the \ac{LCF} approximation agrees within the statistics of the worldline numerics.



\section{Conclusions}

In this paper, we have reviewed the worldline numerics numerical technique with
a focus on computing the effective action of \ac{QED} in
non-homogeneous, cylindrically symmetric magnetic fields. The method
uses a Monte Carlo generated ensemble of worldline loops to
approximate a path integral in the worldline formalism. These
worldline loops are generated using a simple algorithm and encode the
information about the magnetic field by computing the flux through the
loop and the action acquired from transporting a magnetic moment
around the loop.  This technique preserves Lorentz symmetry exactly
and can preserve gauge symmetry up to any required precision.

We have discussed implementing this technique on \ac{GPU} architecture using 
\ac{CUDA}. The main advantage of this architecture is that it allows for a 
very large number of concurrent threads which can be utilized with 
very little overhead. In practice, this means that a large ensemble 
of worldlines can be computed concurrently, thus allowing for a considerable 
speedup over serial implementations, and allowing for the precision of the 
numerics to increase according to the number of threads available. 

This work was supported by the Natural Sciences and Engineering Research
Council of Canada, the Canadian Foundation for Innovation, the British
Columbia Knowledge Development Fund.  It has made used of the NASA ADS
and arXiv.org.
\bibliographystyle{prsty}
\bibliography{references}

\lstset{language=C, otherkeywords={__device__, __global__, float4,
	__launch_bounds__, __CUDA_ARCH__, cudaError_t, define, endif, ifndef, \#}}
\phantomsection	
\addcontentsline{toc}{part}{Appendices}

\appendix*
\section{\texorpdfstring{\acs{CUDA}}{CUDA}fication of Worldline Numerics}
\label{ch:cudafication}

Our implementation of the worldline numerics technique used a ``co-processing"
(or heterogeneous)
approach where the \ac{CPU} and \ac{GPU} are both used in unison to 
compute different aspects of the problem.  The \ac{CPU} was used 
for computing the spatial and proper-time integrals while the 
\ac{GPU} was employed to compute the contributions to the Wilson 
Loop from each worldline in parallel for each value of $\rho_\mathrm{ cm}$ and $T$.
This meant that each time the integrand was to be computed, it could 
be computed thousands of times faster than on a serial implementation.
A diagram illustrating the co-processing approach is shown in figure 
\ref{fig:coprocessing}.

Code which runs on the \ac{GPU} device must be implemented using 
specialized tools such as the 
\ac{CUDA}-C programming language~\cite{cudazone, CUDAGuide3.2}.
The \ac{CUDA} language is an extension of the C programming language
and compiles under a proprietary compiler, nvcc, which is based off 
of the \acs{GNU} C compiler, gcc. 

\subsection{Overview of \texorpdfstring{\acs{CUDA}}{CUDA}}

\ac{CUDA} programs make use of special functions, called kernel
functions, which are executed many times in parallel. Each parallel
thread executes a copy of the kernel function, and is provided with a
unique identification number,
\lstinline{threadIdx}. \lstinline{threadIdx} may be a one, two, or
three-dimensional vector index, allowing the programmer to organize
the threads logically according to the task.

A program may require many thousands of threads, which are organized into a 
series of organizational units called blocks. For example, the Tesla C1060 
\ac{GPU} allows for up to 1024 threads per block. These blocks are further 
organized into a one or two-dimensional structure called the grid. 
\ac{CUDA} allows for communication between threads within a block, 
but each block is required to execute independently of other blocks.

\ac{CUDA} uses a programming model in which the \ac{GPU} and \ac{CPU} 
share processing responsibilities. The \ac{CPU} runs a host process 
which may call different kernel functions many times over 
during its lifetime. When the host process encounters a kernel 
function call, the \ac{CUDA} device takes control of the processing 
by spawning the designated number of threads and blocks to evaluate 
the kernel function many times in parallel. After the kernel function 
has executed, control is returned to the host process which may then 
copy the data from the device to use in further computations.

The \ac{GPU} device has separate memory from that used by the \ac{CPU}. 
In general, data must be copied onto the device before the kernel is executed 
and from the device after the kernel is executed. 
The \ac{GPU} has a memory hierarchy containing
several types of memory which can be utilized by threads. Each thread 
has access to a private local memory. A block of threads may all access 
a shared memory. Finally, there is memory that can be accessed by any 
thread. This includes global memory, constant memory, and texture memory. 
These last three are persistent across kernel launches, meaning that 
data can be copied to the global memory at the beginning of the program 
and it will remain there throughout the execution of the program.

\subsection{Implementing \texorpdfstring{\acs{WLN}}{WLN} 
in \texorpdfstring{\acs{CUDA}}{CUDA}-C}

In the jargon of parallel computing, an embarrassingly parallel problem 
is one that can be easily broken up into separate tasks that do not 
need to communicate with each other. The worldline technique is one 
example of such a problem: the individual contributions from each worldline
can be computed separately, and do not depend on any information from 
other worldlines. 

Because we may use the same ensemble of worldlines throughout 
the entire calculation, the worldlines can be copied into the 
device's global memory at the beginning of the program. The global memory 
is persistent across all future calls to the kernel function.
This helps to reduce overhead compared to 
parallelizing on a cluster where the worldline data would need 
to be copied many times for use by each \ac{CPU}.
The memory copy can be done from 
\ac{CPU} code using the built-in function \lstinline{cudaMemcpy()} and 
the built-in flag \lstinline{cudaMemcpyHostToDevice}.
\begin{widetext}
\begin{lstlisting}
//Copy worldlines to device
errorcode = cudaMemcpy(worldlines_d, worldlines_h,
	nThreads*nBlocks*Nppl*sizeof(*worldlines_h),
	cudaMemcpyHostToDevice);
if(errorcode > 0) printf("cudaMemcpy WLs: 
        \%s\n", cudaGetErrorString(errorcode));
\end{lstlisting}
\end{widetext}
In the above, \lstinline{worldlines_d} and \lstinline{worldlines_h} are 
pointers of type \lstinline{float4} (discussed below) which point to the worldline 
data on the device and host, respectively. \lstinline{nThreads},
\lstinline{nBlocks}, and \lstinline{Nppl} are integers representing the 
number of threads per block, the number of blocks, and the number 
of points per worldline. 
So, \lstinline{nThreads*nBlocks*Nppl*sizeof(*worldlines_h)}
is the total size of memory to be copied. The variable \lstinline{errorcode}
is of a built-in \ac{CUDA} type, \lstinline{cudaError_t} which returns 
an error message through the function \lstinline{cudaGetErrorString(errorcode)}.

The global memory of the device has a very slow bandwidth compared to 
other memory types available on the \ac{CUDA} device. 
If the kernel must access the worldline 
data many times, copying the worldline data needed by the block of threads 
to the shared memory of that block will provide a performance increase. 
If the worldline data is not too large for the device's constant memory, 
this can provide a performance boost as well since the constant memory 
is cached and optimized by the compiler. 
However, these memory optimizations are not used here 
because the worldline data is too large for constant memory and 
is not accessed many times by the kernel. This problem can 
also be overcome by generating the loops on-the-fly directly on the 
\ac{GPU} device itself without 
storing the entire loop in memory~\cite{2011arXiv1110.5936A}.

In order to compute the worldline Wilson loops, we must create 
a kernel function which can be called from \ac{CPU} code, but 
which can be run from the \ac{GPU} device.  Both must 
have access to the function, and this is communicated to the compiler
with the \ac{CUDA} function prefix \lstinline{__global__}.  
\begin{widetext}
\begin{lstlisting}
#define THREADS_PER_BLOCK 256
#if __CUDA_ARCH__ >= 200
    #define MY_KERNEL_MAX_THREADS (2 * THREADS_PER_BLOCK)
    #define MY_KERNEL_MIN_BLOCKS 3
#else
    #define MY_KERNEL_MAX_THREADS (2 * THREADS_PER_BLOCK)
    #define MY_KERNEL_MIN_BLOCKS 2
#endif

__global__ void 
__launch_bounds__(MY_KERNEL_MAX_THREADS, MY_KERNEL_MIN_BLOCKS)
__global__ void ExpectValue(float4 *Wsscal, float4 *Wsferm, 
	float4 *worldlines, float4 xcm, float F, 
	float l2, float rtT, int Nl, int Nppl, int fermion)
//Each thread computes the Wilson loop value for a single 
//worldline identified by inx.
{
        int inx = blockIdx.x * blockDim.x + threadIdx.x;        
        WilsonLoop(worldlines, Wsscal, Wsferm, 
			xcm, inx, F, l2, rtT, Nppl, fermion);
            
}
\end{lstlisting}
\end{widetext}
The preprocessor commands (\ie the lines beginning with \#) 
and the function \lstinline{__launch_bounds__()} provide 
the compiler with information that helps it minimize the registers needed and prevents 
spilling of registers into much slower local memory. More information can be found in 
section B.19 of the \ac{CUDA} programming guide~\cite{CUDAGuide3.2}.
The built-in variables \lstinline{blockIdx}, \lstinline{blockDim}, and \lstinline{threadIdx} 
can be used as above to assign a unique index, \lstinline{inx}, 
to each thread.
\ac{CUDA} contains a native vector data type called \lstinline{float4} 
which contains a four-component vector of data which can be copied 
between host and device memory very efficiently.
This is clearly 
useful when storing coordinates for the worldline points or the 
center of mass.  These coordinates are accessed using 
C's usual structure notation: \lstinline{xcm.x}, \lstinline{xcm.y}, \lstinline{xcm.z}, \lstinline{xcm.w}.
We also make use of this data type to organize the output 
Wilson loop data, \lstinline{Wsscal} and \lstinline{Wsferm}, into the groups discussed in section 
\ref{sec:discunc}.

The function \lstinline{WilsonLoop()} contains code which only the 
device needs access to, and this is denoted to the compiler by the 
\lstinline{__device__} function prefix. For example, 
we have,
\begin{widetext}
\begin{lstlisting}
extern "C"
__device__ void WilsonLoop(float4 *worldlines, float4 *Wsscal, 
	float4 *Wsferm, float4 xcm, int inx, float F, float l2, 
	float rtT, int Nppl, int fermion)
//Compute the Wilson loops for the thread inx and store the 
//results in Wsscal[inx] (scalar part) 
//and Wsferm[inx] (fermion part)
{
	...
}
\end{lstlisting}
\end{widetext}
Note that we pass $\sqrt{T}$ to the function (\lstinline{float rtT}) instead of 
$T$ so that we only compute the square root once instead of once 
per thread.  Avoiding the square root is also why I chose to 
express the field profile in terms of $\rho^2$.

As mentioned above, the \ac{CUDA} 
device is logically divided into groups of 
threads called blocks.  The number of blocks, nBlocks, 
and the number of threads per block, \lstinline{nThreads}, which are to 
be used must be specified when calling \ac{CUDA} kernel functions 
using the triple chevron notation.
\begin{widetext}
\begin{lstlisting}
MyKernel<<<nBlocks,nThreads>>>(void* params)
\end{lstlisting}
\end{widetext}
In the following snippet of code, we call the \ac{CUDA} device 
kernel, \lstinline{ExpectValue()} from a normal C function. 
We then use the \lstinline{cudaMemcpy()} function to copy 
the Wilson loop results stored as an array 
in device memory as \lstinline{params.Wsscal_d}
to the host memory with pointer \lstinline{params.Wsscal_h}.  The 
contents of this variable may then be used by the \ac{CPU} 
using normal C code.
\begin{widetext}
\begin{lstlisting}
//Call to CUDA device
ExpectValue<<<params.nBlocks, params.nThreads>>>(
	params.Wsscal_d, params.Wsferm_d, 
	params.worldlines, params.xcm, params.F, 
	params.l2, rtT,params.Nl, params.Nppl
	);
//Check for errors during kernel execution
errorcode = cudaGetLastError();
if (errorcode > 0) printf(
	"cuda getLastError EV(): %s\n",
	cudaGetErrorString(errorcode)
	);
//Copy device memory back to host
errorcode = cudaMemcpy(
	params.Wsscal_h, params.Wsscal_d,
	params.Nl*sizeof(params.Wsscal_h[0]), 
	cudaMemcpyDeviceToHost
	);
//Check for memory copy errors
if(errorcode > 0) printf(
	"CUDA memcpy scal Error EV(): %s\n",
	cudaGetErrorString(errorcode)
	);	
if(params.fermion == 1) //if fermionic calculation
 {
	//Copy fermion data from device to host
  	errorcode = cudaMemcpy(
		params.Wsferm_h, params.Wsferm_d,
		params.Nl*sizeof(params.Wsferm_h[0]), 
		cudaMemcpyDeviceToHost
		);
	//Check for memory copy errors
	if(errorcode > 0) printf(
		"CUDA memcpy ferm Error EV(): %s\n",
		cudaGetErrorString(errorcode)
		);
 };
\end{lstlisting}
\end{widetext}

\subsection{Compiling \texorpdfstring{\acs{WLN}}{WLN} 
\texorpdfstring{\acs{CUDA}}{CUDA} Code}

Compilation of \ac{CUDA} kernels is done through the Nvidia \ac{CUDA}
compiler driver \lstinline{nvcc}. \lstinline{nvcc} can be provided
with a mixture of host and device code. It will compile the device
code and send the host code to another compiler for processing. On
Linux systems, this compiler is the \acs{GNU} C compiler,
\lstinline{gcc}. In general, \lstinline{nvcc} is designed to mimic the
behaviour of \lstinline{gcc}. So the interface and options will be
familiar to those who have worked with gcc.

There are two dynamic libraries needed for compiling \ac{CUDA}
code. They are called \lstinline{libcuda.so} and
\lstinline{libcudart.so} and are located in the \ac{CUDA} toolkit
install path. Linking with these libraries is handled by the
\lstinline{nvcc} options \lstinline{-lcuda} and
\lstinline{-lcudart}. The directory containing the \ac{CUDA} libraries
must be referenced in the \lstinline{LD_LIBRARY_PATH} environment
variable, or the compiler will produce library not found errors.

Originally, \ac{CUDA} devices did not support double precision
floating point numbers.  These were demoted to \lstinline{float}. More
recent devices do support \lstinline{double}, however. This is
indicated by the compute capability of the device being equal or
greater than 1.3. This capability is turned off by default, and must
be activated by supplying the compiler with the option
\lstinline{-arch sm_13}. The kernel 
primarily uses \lstinline{float} variables because
this reduces the demands on register memory and allows for greater
occupancy.

Another useful compile option is
\lstinline{--ptxas-options="-v"}. This option provides verbose
information about shared memory, constants, and registers used by the
device kernel. The kernel code for the worldline integrals can 
become sufficiently
complex that one may run out of registers on the device.
This can be avoided by paying attention to the number of
registers in use from the verbose compiler 
output, and using the \ac{CUDA} Occupancy Calculator
spreadsheet provided by Nvidia to determine the maximum number of
threads per block that can be supported~\cite{cudaOC}.  The
\ac{CUDA} error checking used in this paper is sufficient to
discover register problems if they arise.
\end{document}